\def\lbol{L$_{\rm Bol}$}
\def\nlyc{N$_{\rm LyC}$}
\def\msun{M$_\odot$}
\def\lsun{L$_\odot$}
\def\h2{H$_2$}
\def\fe2{[\ion{Fe}{2}]}
\def\oa{[\ion{O}{1}]}
\def\ob{[\ion{O}{2}]}
\def\oc{[\ion{O}{3}]}
\def\ne2{[\ion{Ne}{2}]}
\def\s4{[\ion{S}{4}]}

\def\pb{Pa$\beta$}
\def\bd{Br$\delta$}
\def\bg{Br$\gamma$}
\def\ba{Br$\alpha$}
\def\av{$A_{\rm V}$}
\def\gal{NGC~253}
\def\etal{et~al.}
\def\kms{km~s$^{-1}$}
\def\cgs{${\rm erg\ s^{-1}\ cm^{-2}}$}
\def\rah{$\rm^h$}
\def\ram{$\rm^m$}

\documentstyle[11pt,aaspp4]{article}
\begin{document}

\title{The Nuclear Starburst in \gal}
\author{C. W. Engelbracht, M. J. Rieke, G. H. Rieke}
\affil{Steward Observatory, University of Arizona, Tucson, AZ 85721}
\author{D. M. Kelly}
\affil{Wyoming Infrared Observatory, University of
Wyoming, Laramie, WY 82071}
\and
\author{J. M. Achtermann\altaffilmark{1}$^,$\altaffilmark{2}}
\affil{Department of Astronomy/McDonald Observatory, University of
Texas, Austin, TX 78712}
\altaffiltext{1}{Visiting Astronomer at the Infrared Telescope
Facility, which is operated by the University of Hawaii under contract
to the National Aeronautics and Space Administration.}
\altaffiltext{2}{Current address:  Tivoli Systems, Stonelake 6, 11000
Mopac Expressway, Suite 400, Austin, TX 78759}

\begin{abstract}

We have obtained long-slit spectra of \gal\ in the J, H, K, and N
bands, broadband images in the J, H, and K$_s$ bands,
narrowband images centered at the wavelengths of \bg\ and
\h2(1,0)S(1), and imaging spectroscopy centered on
\ne2(12.8\micron).  We have subtracted a composite stellar spectrum
from the galaxy spectrum to measure faint emission lines which
otherwise would be buried in the complicated continuum structure.

We use these data and data from the literature in a comprehensive
re-assessment of the starburst in this galaxy. We confirm that the
\fe2\ emission is predominantly excited by supernova explosions and
show that the rate of these events can be derived from the strength of
the infrared \fe2\ lines. Although the \h2\ emission superficially
resembles a thermally excited spectrum, most of the \h2\ infrared
luminosity is excited by fluorescence in low density gas. We confirm
the presence of a bar and also show that this galaxy has a
circumnuclear ring. The relation of these features to the gaseous bar
seen in CO is in agreement with the general theoretical picture of how
gas can be concentrated into galaxy centers by bars. We derive a
strong upper limit of $\sim$ 37,000K for the stars exciting the
emission lines. We use velocity-resolved infrared spectra to determine
the mass in the starburst region. Most of this mass appears to be
locked up in the old, pre-existing stellar population.  Using these
constraints and others to build an evolutionary synthesis model, we
find that the IMF originally derived to fit the starburst in M~82 also
accounts for the properties of \gal; this IMF is similar to a modified
Salpeter IMF.  The models indicate that rapid massive star formation
has been ongoing for 20-30 million years in \gal---that is, it is in a
late phase of its starburst. Its optical spectrum has characteristics
of a transitional HII/weak-\oa\ LINER. We model the emission line
spectrum expected from a late phase starburst and demonstrate that it
reproduces these characteristics.

\end{abstract}


\section{Introduction}

A significant fraction of nearby galaxies have strong, centrally
concentrated infrared sources indicative of rapid star formation in
their central few hundred parsecs (Rieke and Lebofsky 1979; Devereux,
Becklin, \& Scoville 1987).  These so-called starburst galaxies make a
dramatic contribution to the star formation activity in the local
universe; Heckman (1997) calculates that just four galaxies (of which
\gal\ is one) are responsible for 25\% of the high-mass star formation
within 10~Mpc.  Local starburst galaxies serve as a model for galaxy
formation and evolution at higher redshifts.  An understanding of how
these events occur and evolve will require study over the full range
of scale and age.  At the same time, very detailed modeling is
required of the nearest starbursts such as \gal\ to provide templates
for models of less accessible systems.

\gal\ has been observed extensively in all accessible spectral
regions (e.g., radio: Ulvestad \& Antonucci 1997; molecular line
emission: Mauersberger et al. 1996; far infrared: Carral et al. 1994;
mid-IR: Keto et al.  1994, Telesco, Dressel, \& Wolstencroft 1993;
near-IR: Sams et al. 1994 ; optical: Watson et al. 1996; X-ray: Ptak
et al. 1997; and gamma ray: Paglione et al. 1996). Additional
references are given in those cited above.  \gal\ was one of the
first galaxies to be studied with an evolutionary starburst model
(Rieke et al. 1980). The subsequent observations taken together have
substantially expanded on our knowledge of the starburst phenomenon
from the inputs to that model, demonstrating the presence of
supernovae remnants and HII regions, young ultracompact star clusters,
a hot superwind, and a host of other starburst phenomena.

Young starbursts, including \gal, are characterized by high levels
of interstellar extinction that can make properties deduced from
optical measurements unrepresentative (e.g., Rieke \etal\ 1980; Puxley
1991).  Consequently, infrared measurements play a critical role in
defining nebular properties, morphologies, rotation curves, and other
parameters. We have added to the extensive infrared observations of
this galaxy with the deep NIR images covering a wide ($\sim7^\prime$)
field, narrowband images in \bg\ and \h2, near- and mid-infrared
spectra, and a high-resolution spectral map of \ne2.

Since the first models of M~82 and \gal\ (Rieke et al. 1980),
evolutionary synthesis modeling has become an important component of
starburst study (e.g., Doyon, Joseph, \& Wright 1994; Leitherer \&
Heckman 1995 and references therein; Schinnerer \etal\ 1997).  We
interpret the new infrared measurements with the other properties of
\gal\ through a modern synthesis model (Rieke \etal\ 1993, hereafter
RLRT93). In this way, the behavior of \gal\ can be related
consistently to that of other galaxies we have modelled as part of a
comprehensive study of nearby starbursts: NGC~6946 (Engelbracht \etal\
1996, hereafter ERRL96) and M~82 (McLeod \etal\ (1993); RLRT93); a
number of blue dwarf galaxies (Vanzi \& Rieke 1997), and a sample of
interacting galaxies (Vanzi et al. 1998). We demonstrate that the
\gal\ starburst is at a relatively late stage of development, during
which the high rate of supernova explosions leads to behavior that
places it as a transition object between pure starbursts and weak-\oa\
LINERs.

This paper is organized as follows: In \S\ref{sec:obs} we describe the
observations and the processing we performed on the data.  We then use
our data and results from the literature to derive physical properties
of the starburst in \S\ref{sec:parms}.  In \S\ref{sec:models} we use
the properties derived in the previous section to constrain
evolutionary synthesis models of the starburst.  We use the models to
explore burst parameters such as age, intensity, IMF, and star
formation history.  \S\ref{sec:other} discusses other results obtained
from our data or implied by the starburst modeling results.  In
\S\ref{sec:conclude} we summarize the paper.

\section{Observations and Data Reduction}
\label{sec:obs}

\subsection{Images}

Broadband images in the J, H, and K$_s$ bands as well as narrowband
images with 0.5\% bandwidth filters tuned to \bg\ and the (1,0)S(1)
line of \h2\ were obtained at the Steward Observatory 1.55m telescope
using a camera equipped with a NICMOS3 array.  The data were reduced
as described in ERRL96. The deep broadband images were flux-calibrated
using the UKIRT faint standard number 2 (Courteau 1994). We calibrated
the K$_s$-band image to K-band magnitudes using the standard-star
photometry, and will hereafter refer to this image as the K-band
image.  We have ignored the small color correction between the K$_s$
and K band, since this correction is smaller than the photometric
uncertainty.

We present photometry of the nucleus in
Table~\ref{tbl:phot} in several circular apertures, plus a region the
size of our extracted slit aperture (used to calibrate our spectra.)
Photometric errors were estimated from the dispersion in multiple
measurements of the standard star at different airmasses and are
typically 2\% in the J and H bands and 4\%-5\% in the K band.

\begin{figure}
\plotone{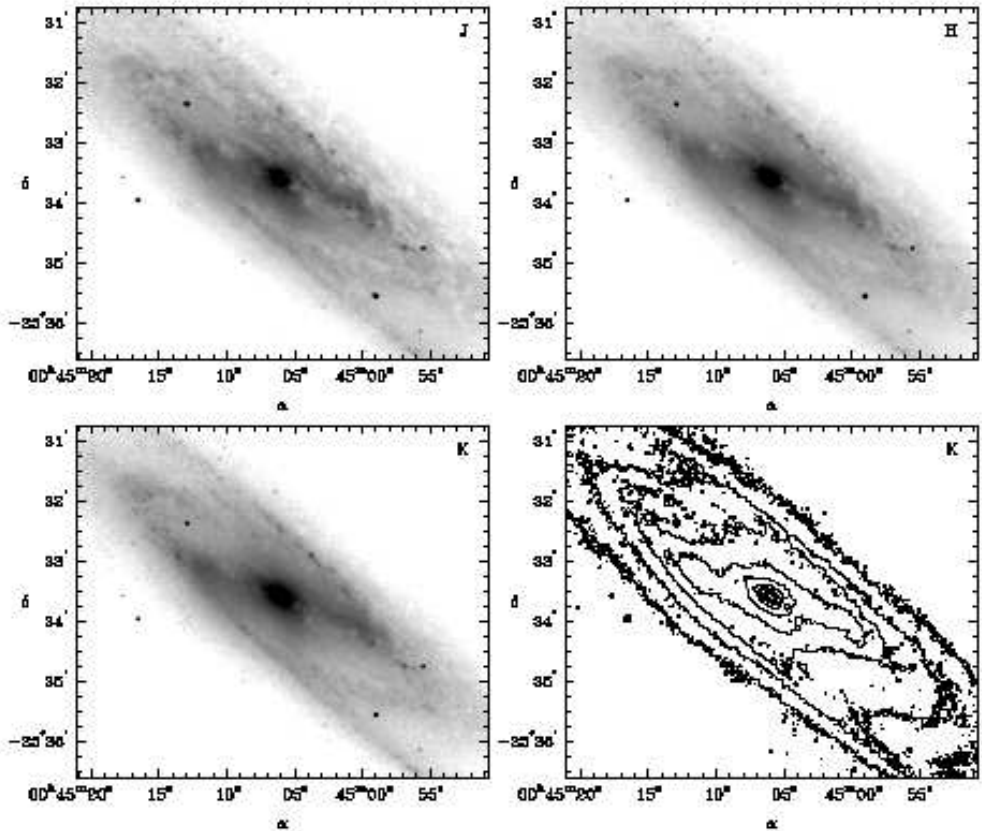}
\caption{Greyscale images of \gal\ in J, H, and K bands, plus a
contour map at K.  The display ranges are 17 to 20, 16.2 to 19.2, 16
to 19, and 14 to 19.4 magnitudes in the J, H, K, and K-contour images,
respectively.}
\label{fig:bbimages}
\end{figure}

The broadband images are presented in Figure~\ref{fig:bbimages}.  The
images have a scale of 0\farcs9 per pixel and the coordinate system is
plotted assuming that the intensity peak of the infrared images
corresponds to the position of \gal\ quoted in NED (Nasa Extragalactic
Database): $\alpha (1950) = 00$\rah\ 45\ram\ 05\fs9, $\delta (1950) =
-25$\arcdeg\ 33\arcmin\ 40\arcsec.  This position closely matches the
position of the K-band peak found by Sams \etal\ (1994).

\begin{figure}
\plotone{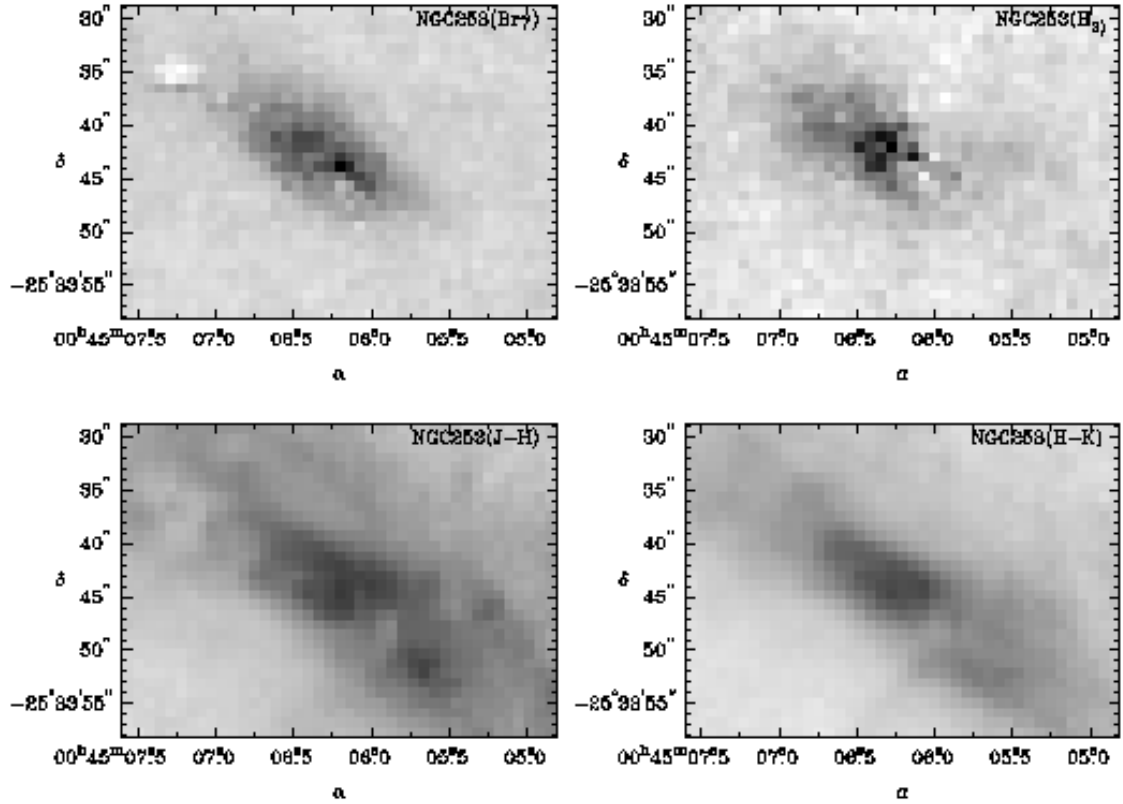}
\caption{Greyscale images of the nuclear region of \gal\ in narrow,
continuum-subtracted bands centered at the lines of \h2 (1,0)S(1)
and \bg, plus $J-H$ and $H-K$ color maps of the same region.  The
display range is 0.6 to 1.5 magnitudes for the $J-H$ image and 0.1 to
1.2 magnitudes for the $H-K$ image.}
\label{fig:nbimages}
\end{figure}

The narrow-band images are presented in Figure~\ref{fig:nbimages}.
These images are the differences between the on-line and off-line
images.  In the same figure, we present $J-H$ and $H-K$ color maps of
the nuclear region, which show that the galaxy colors become very red
on the nucleus due to strong extinction.

\subsection{Near-IR spectra}

We obtained long-slit spectra of \gal\ in the J, H, and K bands at a
resolution of $\sim3000$ with the Steward Observatory 2.3m telescope
using FSpec, a near-infrared long-slit spectrometer (Williams \etal\
1993).  The slit was centered on the infrared peak of the galaxy with
a slit-viewing infrared guide camera.  Observing and data-reduction
techniques are as described in ERRL96, with the exception that a more
precise wavelength calibration was obtained using a set of numerically
computed wavelengths for the OH airglow lines (C. Kulesa, private
communication).

To correct the spectra for atmospheric absorptions, a spectrum of
HR~173, a G3V star about 2\arcdeg\ away from \gal, was taken close in
time to the galaxy observations and was divided into the galaxy
spectrum.  This step introduced spurious emission features into the
spectrum due to intrinsic absorptions in the standard star.  To
correct for these features, we multiplied our galaxy spectrum by the
high-resolution solar spectrum (smoothed to our instrumental
resolution) obtained by Livingston \& Wallace (1992).

\begin{figure}
\plotone{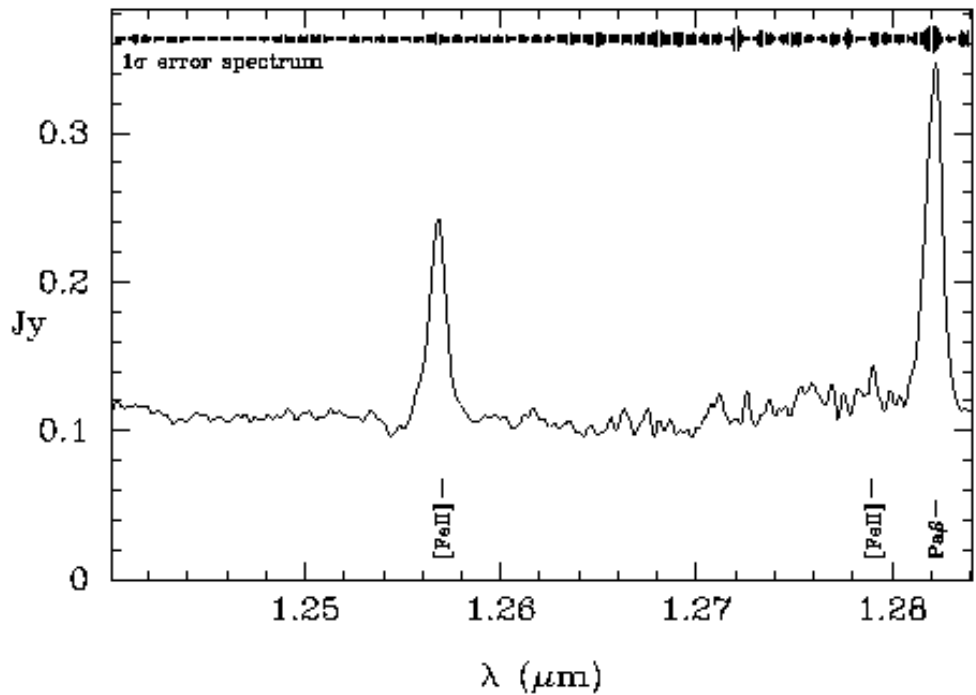}
\caption{High-resolution J-band spectrum of \gal.  The slit was
oriented along the plane of the galaxy and was 2\farcs4 wide.  The
spectrum shown here is the sum of the central 12\arcsec.  The spectrum
was flux-calibrated as described in the text and has been shifted to
zero velocity.  Above the flux spectrum is plotted the 1-sigma error
spectrum.}
\label{fig:jspec}
\end{figure}

\begin{figure}
\plotone{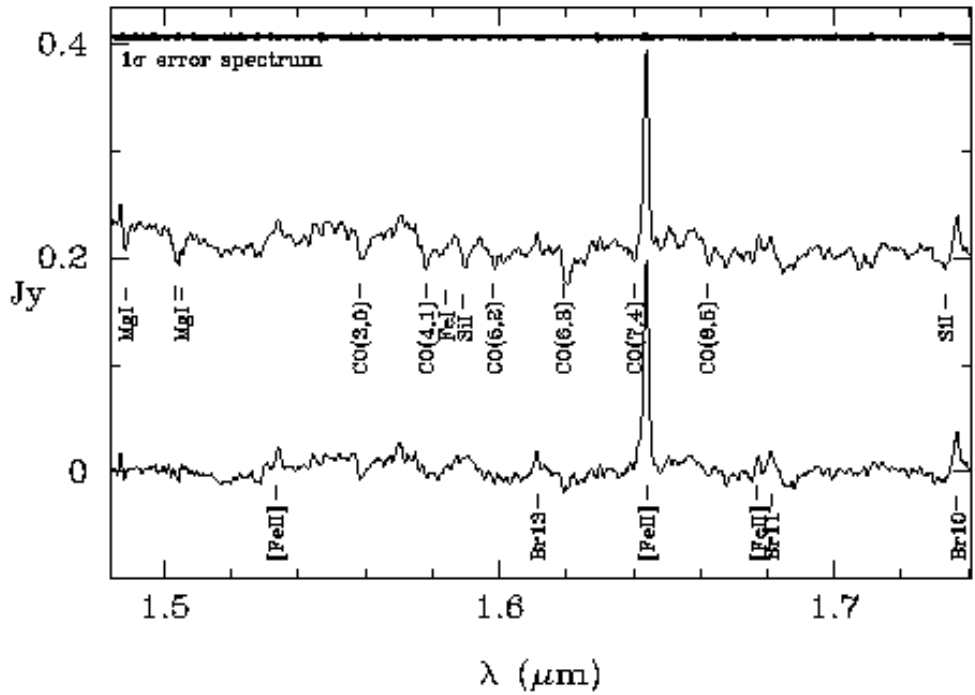}
\caption{High-resolution H-band spectrum of \gal.  Details as for the
J-band spectrum.  The lower spectrum is the result of subtracting a
stellar continuum spectrum (produced by combining several late-type
stellar spectra as described in the text) from the galaxy spectrum.}
\label{fig:hspec}
\end{figure}

\begin{figure}
\plotone{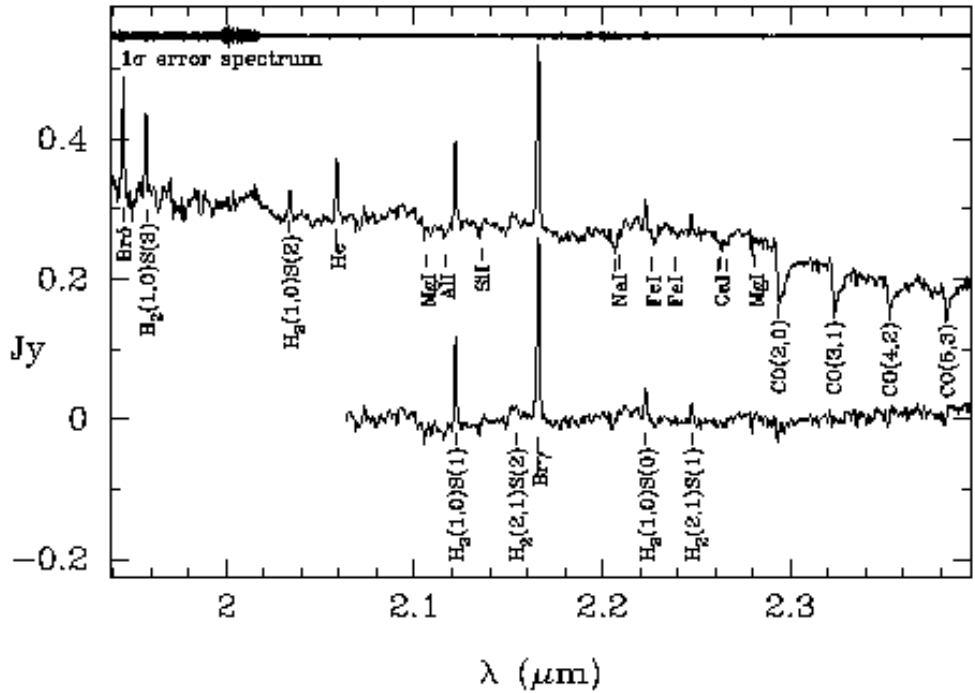}
\caption{High-resolution K-band spectrum of \gal.  Details as for
J-band spectrum, and a stellar continuum subtraction has been
performed as for the H-band spectrum.  The region shortward of
2.06\micron\ contains few significant stellar absorptions and so many
of our stellar spectra do not cover this region.  Note that the error
spectrum indicates the spectrum is significantly noisier shortward of
2.02\micron, where the atmospheric transmission becomes poor.}
\label{fig:kspec}
\end{figure}

Most of the spectra were obtained with the slit oriented along the
major axis of the galaxy, while two key grating settings which cover
the (1,0)S(1) line of \h2, \bg, and the (2,0) and (3,1) CO bands were
obtained with the slit oriented along the minor axis. The J, H, and
K-band spectra are presented in Figures \ref{fig:jspec},
\ref{fig:hspec}, and \ref{fig:kspec}, respectively.  Each spectrum is
the sum of 10 1\farcs2 pixels along the slit, centered on the
intensity peak.  We have computed a 1-$\sigma$ uncertainty for each
point in the spectrum from the dispersion in the set of individual
observations.  This error spectrum is plotted above each flux
spectrum.  The flux scale for each spectrum was taken from the
broadband images, using the 2\farcs4$\times$12\arcsec\ aperture in
Table~\ref{tbl:phot}.

We obtained additional spectra of \gal\ in the J and K bands at a
resolution of $\sim$ 1200 using the same instrument and telescope.
The slit was again oriented along the major axis of the galaxy and we
extracted a similar aperture from these spectra.  The spectra were
reduced in a similar fashion to the high-resolution spectra, with the
exception that our standard star was HR~232, an A3V, so the only
correction for intrinsic absorption was to interpolate over the H
lines.  The use of a hotter standard star (in which metal lines are
presumably weak or undetectable at this resolution) to correct these
spectra allows us to confirm that our correction for the intrinsic
absorptions (other than H) in the standard star used for the
high-resolution data was accurate.

\begin{figure}
\plotone{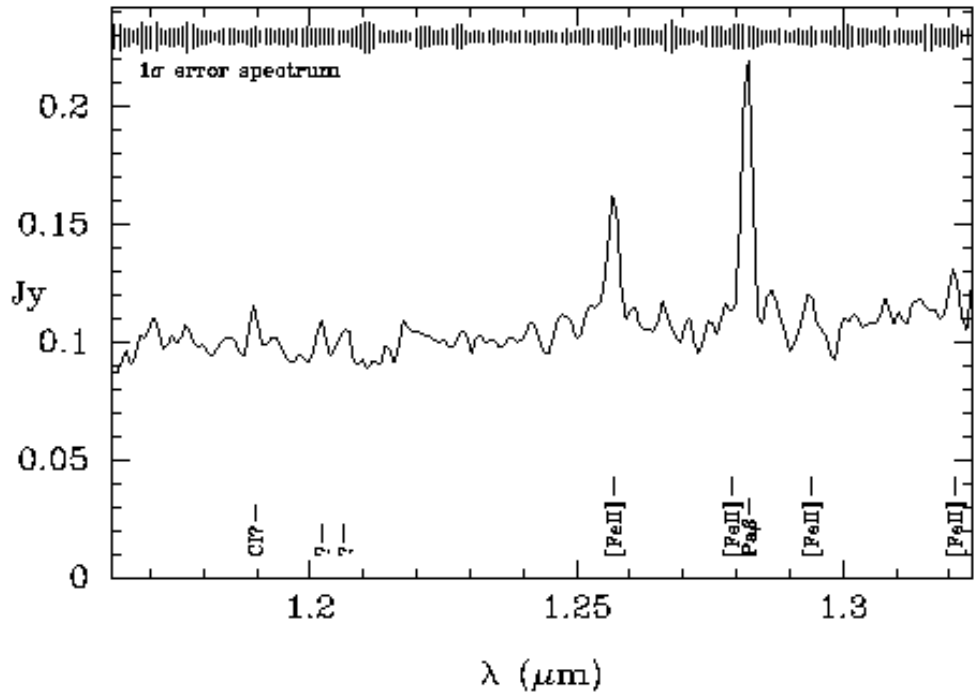}
\caption{Low-resolution J-band spectrum of \gal.  Details as for the
high-resolution spectrum.}
\label{fig:jspec-lr}
\end{figure}

\begin{figure}
\plotone{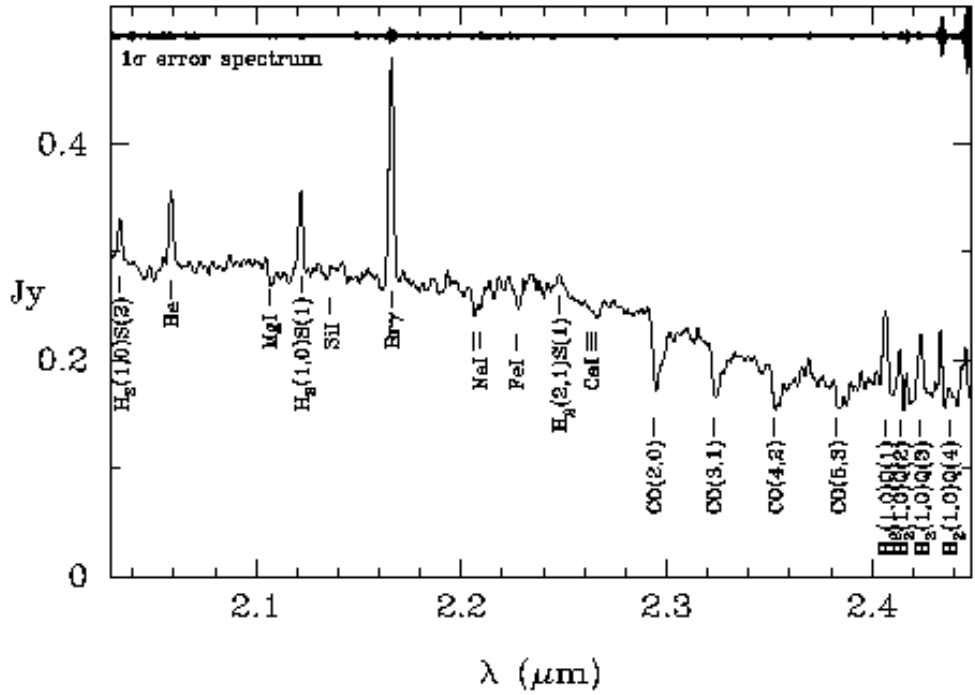}
\caption{Low-resolution K-band spectrum of \gal.  Details as for
J-band spectrum.  The apparent emission features beyond 2.4\micron\
that are not labeled as \h2\ lines correspond to regions of poor
atmospheric transmission, as indicated by the spikes in the error
spectrum at those points.}
\label{fig:kspec-lr}
\end{figure}

The low-resolution spectra (Figures~\ref{fig:jspec-lr} and
\ref{fig:kspec-lr}), while lacking some of the detail in the
high-resolution ones, give access to several additional \fe2\ and \h2\
lines as well as some unidentified features in the J band. The strong
features beyond 2.4\micron\ not marked as \h2\ lines are narrow
telluric absorptions which are poorly sampled at our low resolution
and which appear as emission features after the spectrum is divided by
the standard star. All measurements taken from the near-infrared
spectra are derived from the high-resolution spectra when possible.

\subsection{Faint emission lines}

A difficulty with interpreting near-infrared spectra of starburst
galaxies is that weak nebular emission lines are lost in the structure
of the stellar absorption features. To counteract this problem, we
have used a library of stellar spectra (to be published separately) to
model the galaxy continuum.  We selected a suite of fifteen stars
which minimized the residual continuum when subtracted from the galaxy
spectrum.  The resulting mix of stars is thus not optimized in a
population synthesis sense, but was empirically chosen to be a good
match to the galaxy continuum.  We redshifted the stellar spectra to
the galaxy velocity, broadened them to match the velocity dispersion
in the galaxy (see \S{\ref{sec:disp}}), scaled them, and subtracted
them from the galaxy spectrum.  The continuum-subtracted spectra of
\gal\ are shown in the lower parts of Figures~\ref{fig:hspec} and
\ref{fig:kspec}.  The H-band and K-band flux measurements of emission
lines presented in this paper are taken from the continuum-subtracted
spectra.

\subsection {Mid-IR Spectra}
\label{sec:midir}

We obtained spectra of the \s4\ 10.5 \micron\ line in \gal\ using the
Irshell spectrograph (Lacy et al.~1989) on the NASA Infrared Telescope
Facility.  The 10\arcsec\ long by 2\arcsec\ wide slit was positioned
across the nucleus of the galaxy at PA=61\arcdeg.  Additional spectra
were measured with parallel slit positions, one slit width to either
side of the nucleus.  The 10$\times$64 Si:As array provided 1\arcsec\
sampling along the slit and a spectral resolution of 30~\kms.  The
seeing was poor and the sky was partly cloudy.  DC sky levels were
monitored to identify frames that were contaminated by clouds.  A
dome-temperature card, chopped against the sky, was used for
flat-fielding, atmospheric correction, and fluxing (Lacy et al.~1989).
The data were reduced using a software package developed by Achtermann
(1994).

\begin{figure}
\plotone{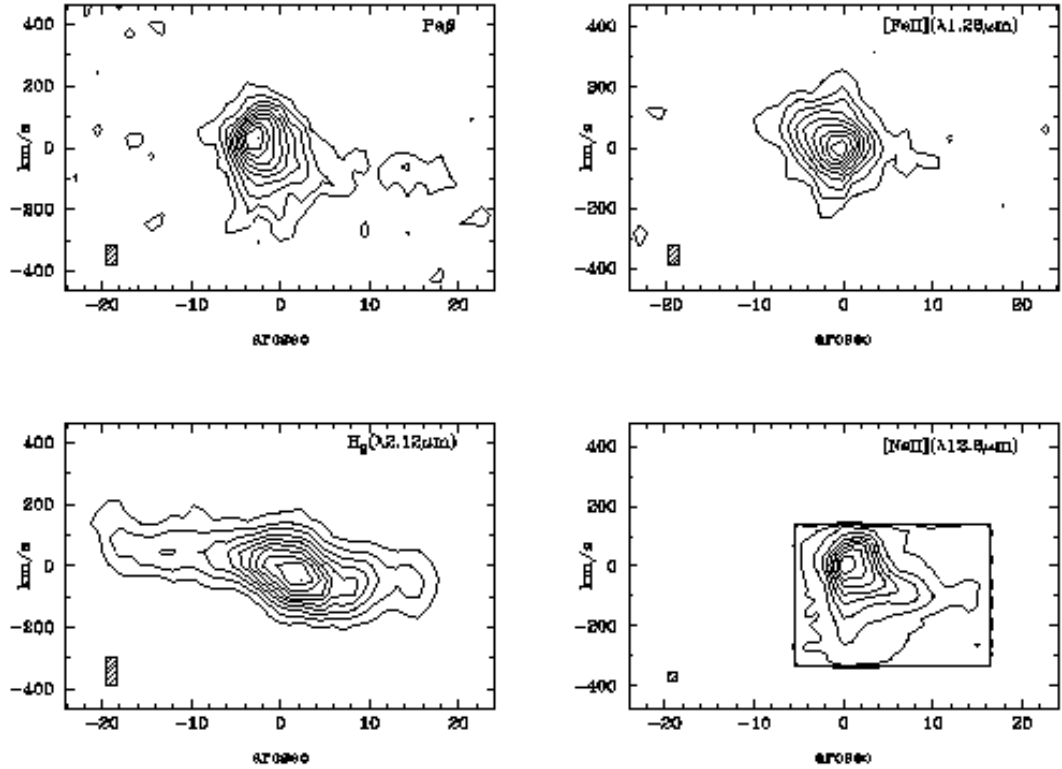}
\caption{Position-velocity plots along the major axis for several
strong lines.  NE is on the right and SW on the left.  In each case,
the upper 90\% of the contours are plotted.  The hatched region in the
lower-left-hand corner of each plot indicates the size of a resolution
element along the spatial and dispersion axes.  The \ne2\ observations
were obtained in the 22\arcsec\ boxed region shown on the plot.}
\label{fig:pv}
\end{figure}

We mapped the \ne2 12.8\micron\ emission from \gal\ by using Irshell
in scanning mode (Lacy \& Achtermann 1994), sweeping the slit back and
forth across the galaxy by 22\arcsec.  The off-source positions at the
ends of the scans were averaged to produce a sky frame. The data were
combined to form a 22\arcsec$\times$22\arcsec\ by 64 spectral channel
data cube.  A position-velocity plot of the \ne2\ line is shown in
Figure~\ref{fig:pv}.  The \ne2\ has a spatial extent of 6\arcsec\ and
a velocity width of 150~\kms\ FWHM.  The integrated
\ne2\ flux is 4.6$\pm$0.1 $\times$ 10$^{-11}$~\cgs. For comparison,
Roche \& Aitken (1985) report a measurement of 3.3 $\times$
10$^{-11}$~\cgs.

\begin{figure}
\plotone{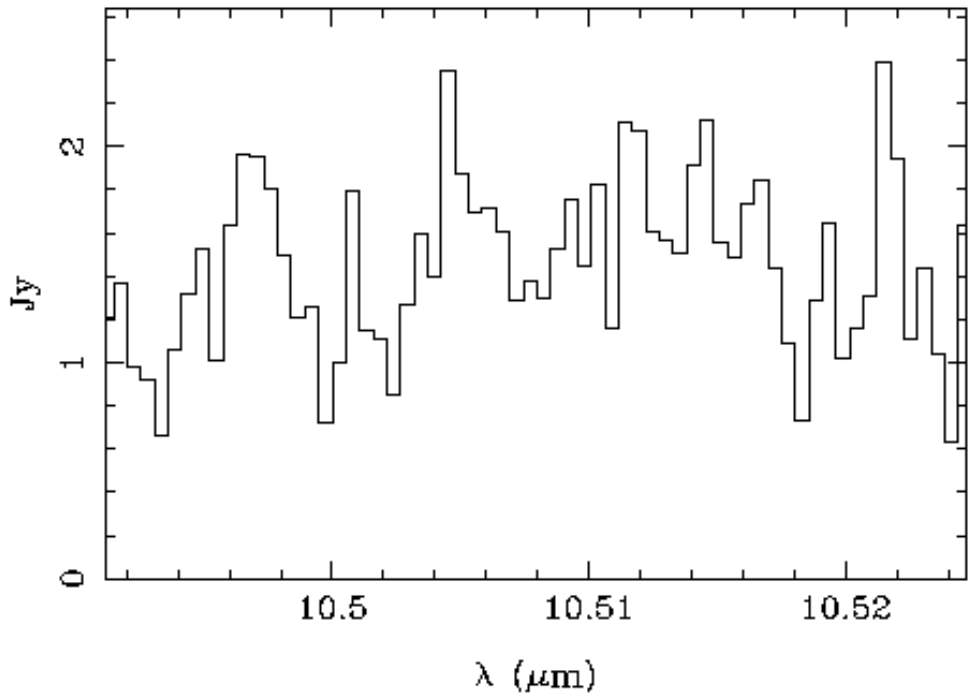}
\caption{MIR spectrum of the central region of \gal, used to determine
an upper limit to the [\ion{S}{4}] flux.  See details in
Section~\ref{sec:midir}.}
\label{fig:midir}
\end{figure}

The \s4\ spectra showed a compact (2\arcsec$\times$2\arcsec) continuum
source with only a hint of line emission at a level of
$5.5\times10^{-14}$~\cgs.  Since the \ne2\ is spread over 6 spatial
and 10 spectral pixels, we determined an upper limit of $3.0 \times
10^{-13}$~\cgs\ on the \s4\ flux by summing over 6 spatial rows (see
Figure~\ref{fig:midir}) and 10 spectral pixels.

\section{Starburst Parameters}
\label{sec:parms}

The values of the parameters that characterize the \gal\ starburst 
are summarized in Table~\ref{tbl:summ}. They have been determined 
assuming a distance of 2.5 Mpc (de Vaucouleurs 1978), at the 
low end of the range of 2.5-3.5 Mpc in the literature. As discussed 
in ERRL96, a small distance places the most conservative constraints on our
starburst models.

The reference aperture we will use is 7\farcs5 in radius.  We chose
this radius because most of the \ne2, \bg, and \h2\ flux (as measured
in our spectra and narrow-band images) and the radio flux (as measured
in the continuum observations of Antonucci \& Ulvestad (1988) or
Turner \& Ho (1983)) arise from within it.  The 10, 20 and 30\micron\
maps of \gal\ from Telesco, Dressel, \& Wolstencroft (1993) as well as
the far-infrared measurements by Smith \& Harvey (1996) also show that
most of the mid- and far-infrared flux is produced in this region.

\subsection{Extinction}
\label{sec:extinct}

We can use line ratios from our spectra to determine the extinction.
If the hydrogen recombination line ratios from Table~\ref{tbl:hlines}
obey Case B, a foreground screen model with \av$=9.2\pm0.7$ can fit
them (c.f. Calzetti, Kinney, \& Storchi-Bergmann 1996). (We have used
an analytical fit to the Rieke \& Lebofsky 1985 extinction law, the
Hummer \& Storey 1987 theoretical recombination line ratios, and have
included an 8\% uncertainty in the relative flux calibration in the
line ratio uncertainties.) This fit roughly agrees with the \ba\
strength (Beck \& Beckwith 1984). However, the mm-wave H recombination
lines still suggest that a significant amount of ionized gas may be
hidden by extreme optical depths (Puxley et al. 1997). The ratio of
\fe2(1.257\micron) to \fe2(1.644\micron) should be 1.35
according to the A-values of Nussbaumer \& Storey (1988), although
this value is subject to some uncertainty.  The observed value is
$0.68\pm0.07$ (Table~\ref{tbl:felines}), which implies
\av$=8.42\pm1.26$, consistent with the estimate from the H
recombination lines.

The colors of our broadband images can also be used to estimate
extinction.  A typical evolved stellar population has $J-H$ and $H-K$
colors of 0.7 and 0.2 (Aaronson 1977), and our starburst models
(RLRT93) show that the same is true for the starburst population. Our
spectra show that the near-infrared continuum is dominated by
starlight, as the strengths of the absorption features are typical of
a population of supergiants and are consistent between the H and K
bands, thus arguing against veiling by a featureless
continuum. Therefore, we can assume that colors redder than nominal
are purely due to reddening.

From Table~\ref{tbl:phot}, the near-infrared colors in our 15\arcsec\
reference aperture are $J-H=1.09$ and $H-K=0.66$, implying color
excesses of 0.39 and 0.46, respectively. Using an analytical fit to
the Rieke \& Lebofsky (1985) extinction law and a foreground screen
geometry for the dust, $E(J-H) = 0.39$ corresponds to $4.26\pm0.91$
magnitudes of visual extinction while $E(H-K) = 0.46$ implies
\av\ $= 7.35\pm1.33$. The extinction values in the near-infrared bands
implied by our $H-K$ color are $A_J=2.00\pm0.36$, $A_H=1.33\pm0.24$,
and $A_K=0.87\pm0.16$. The extinction to the stars appears to be
slightly smaller (at $\sim$3$\sigma$ significance) than that to the
ionized region.

If we adopt a more realistic dust geometry than a foreground screen,
such as the "starburst" model of Witt, Thronson, \& Capuano (1992), we
obtain a better fit to the observed colors. The colors in the nuclear
region of \gal\ are slightly redder than the most heavily extinguished
models presented by Witt \etal, so we have used polynomial fits to
extrapolate beyond their values.  The models imply a total extinction
to the center of the system, \av (cen), of $17.27\pm 6.78$ and
$19.07\pm 3.32$, based on the $J-H$ and $H-K$ excesses respectively.
Although this model contains much more total dust than a foreground
screen model, much of the light comes from relatively unextinguished
stars, so that the implied average equivalent extinction is similar to
the foreground screen case.  The equivalent extinction values in the
near-infrared are $A_J=1.64\pm0.51$, $A_H=1.24\pm0.43$, and
$A_K=0.85\pm0.30$ from the $J-H$ color and $A_J=1.84\pm0.32$,
$A_H=1.40\pm0.24$, and $A_K=0.94\pm0.16$ from the $H-K$ color.

Since the equivalent extinction values in the near infrared are
similar for the two dust geometries, the near infrared fluxes can be
corrected reliably. We will adopt extinction values that are an
average of those derived from the mixed model: $A_K=0.90\pm0.17$,
$A_H=1.32\pm0.25$, and $A_J=1.74\pm0.30$.

\begin{figure}
\plotone{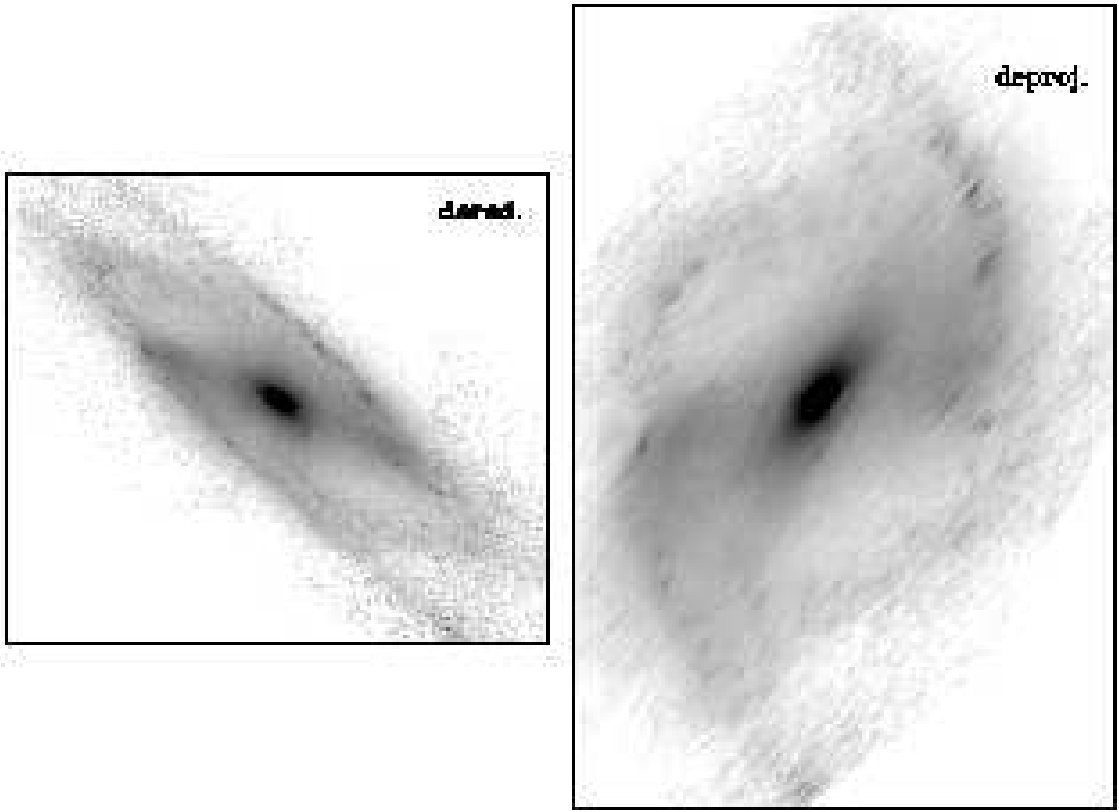}
\caption{K-band image of \gal\ dereddened as described in the text,
then deprojected along the minor axis to determine how the galaxy
would appear if viewed face on.  The images are presented on a
logarithmic intensity scale.  The scale and orientation of the images
are the same as Figure~\ref{fig:bbimages}.}
\label{fig:deproj}
\end{figure}

We have used the $H-K$ color map to deredden our K-band image.  We
assumed that the intrinsic $H-K$ color of the galaxy was 0.2 (so that
$E(H-K) = (H-K) - 0.2$) and converted the color excess to a
dereddening map using the Rieke \& Lebofsky (1985) extinction law. The
K-band image was then multiplied by the dereddening image to produce
the extinction-corrected K-band image of the galaxy presented in
Figure~\ref{fig:deproj}.  We also deprojected the dereddened image by
stretching it along the minor axis by a factor $1/{\rm cos}(i)$, where
we have taken $i=78\arcdeg$ (Pence 1981).  The resulting image is also
presented in Figure~\ref{fig:deproj}.

\subsection{\fe2\ emission}
\label{sec:fe2}

\begin{figure}
\plotone{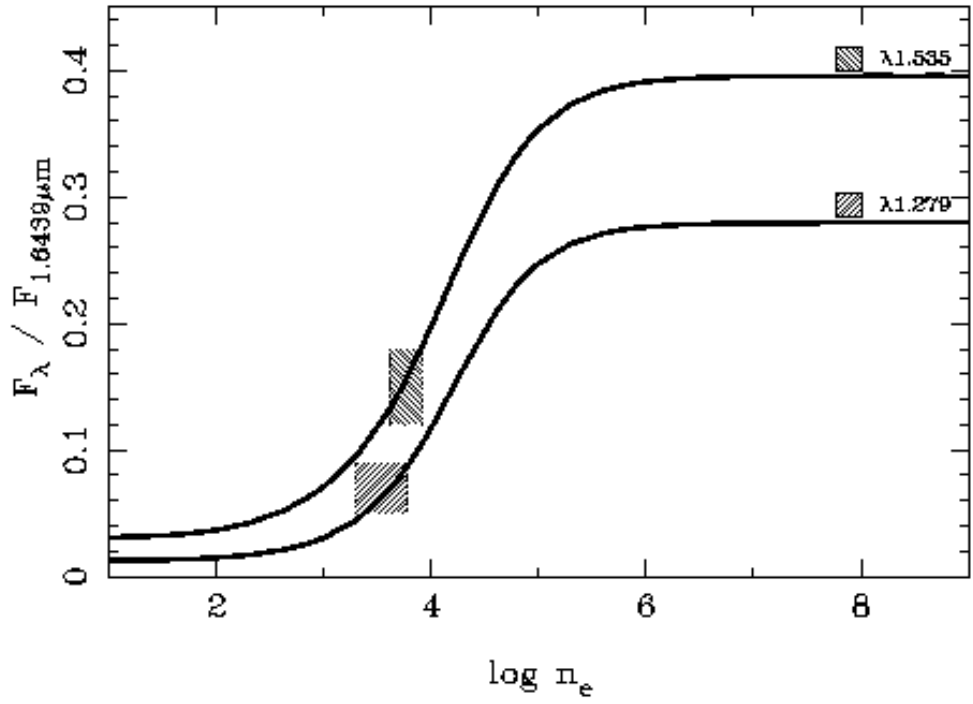}
\caption{Plot of \fe2\ line ratios as a function of density.  The line
ratios have been corrected for the extinction derived in
\S\ref{sec:extinct}.  The range allowed by the observations is
indicated by the shaded regions. The diagnostic lines we have used
here indicate that log(n$_e$) in the \fe2 -emitting region is
$3.7\pm0.13$.}
\label{fig:feplot}
\end{figure}

The high sensitivity and broad wavelength coverage of our
near-infrared spectra has allowed us to measure seven \fe2\ lines,
listed in Table~\ref{tbl:felines}, allowing an estimate of the
electron density in the \fe2-emitting region. We use the ratios
$\lambda1.279$\micron/$\lambda1.644$\micron\ and
$\lambda1.533$\micron/$\lambda1.644$\micron\ and the models of
Bautista \& Pradhan (1996). In the range of interest, the diagnostic
ratios we have chosen are essentially independent of temperature.  We
have plotted the values of these two line ratios (corrected for
extinction) as a function of density in Figure~\ref{fig:feplot}.  For
each curve, we have marked the range consistent with the observations
as a shaded box, indicating a mean density of $\sim5000\ {\rm
cm}^{-3}$.

\fe2\ emission in starbursts is thought to arise predominantly from
supernova remnants. For example, Greenhouse \etal\ (1991)
find peaks on bright radio compact sources in the \fe2\ surface
brightness in M~82. However, these compact sources account for only a
small fraction of the total \fe2\ emission. Forbes \& Ward (1993) find
a correlation between nonthermal 6cm emission and
\fe2. However, the trend line is offset from that for supernova
remnants. Vanzi \& Rieke (1997) explain this offset in terms of the
short lifetime for the \fe2-emitting stage in a supernova remnant.
Our data add to the arguments favoring a supernova origin for the
\fe2, since the density we derive is in good agreement with that
observed in the supernova remnants observed by Oliva, Moorwood, \&
Danziger (1989).

Vanzi \& Rieke (1997) used the observed supernova rate in M~82 to
derive a relation between the \fe2\ luminosity and the supernova rate.
To apply this relation, we correct our observed \fe2\
($\lambda$1.644\micron) flux (from Table~\ref{tbl:felines}) to a
15\arcsec\ aperture using the ratio of the K-band fluxes in the
15\arcsec\ and 2\farcs4~$\times$~12\arcsec\ apertures (this scaling is
reasonable because the K-band flux is dominated by cool supergiants
which are the immediate precursors to the supernova remnants).  We
also correct for 1.33 magnitudes of extinction. We obtain a total
\fe2\ flux of $3.8\times10^{-12}$~\cgs, or a luminosity of
$2.8\times10^{39}$~erg/s. The Vanzi \& Rieke relation then implies a
supernova rate of 0.03 per year.  If the \fe2-emitting phase of each
supernova remnant lasts $\sim10^4$ years, this rate implies that there
are currently $\sim300$ \fe2-emitting supernova remnants in the
nucleus of \gal.

Ulvestad \& Antonucci (1997) find an upper limit on the supernova rate
of 0.3~yr$^{-1}$.  The supernovae have also been modeled by Colina \&
P\'{e}rez-Olea (1992), who estimate a rate of 0.05 yr$^{-1}$, and by
Van Buren \& Greenhouse (1994) and Paglione et al. (1996), who both
find 0.08 yr$^{-1}$.  These estimates are all based on modeling
supernova counts and hence are to first order
distance-independent. The SN rate can also be estimated from other
luminosities than that of \fe2.  For example, if we assume the
nonthermal radio flux from Turner \& Ho (1983) is entirely due to
synchrotron emission from supernova remnants, we can derive a
supernova rate of 0.03~yr$^{-1}$ using equation (8) of Condon \& Yin
(1990).

All of the other rate estimates are in satisfactory agreement with our
estimate using the \fe2\ emission. Both of the luminosity-based
estimates are lower than those from supernova counts by a factor of
about two. Although this difference is probably within the errors, if
the galaxy were at 3.5~Mpc (within the uncertainty in distance), the
luminosity-based rates would double and the agreement would be
improved.

\subsection{Temperature of the Hot Stars}
\label{sec:temp}

The difference in ionization potentials of S$^{++}$ and Ne$^0$ makes
the \s4/\ne2\ ratio a sensitive indicator of the effective temperature
of the UV radiation field. The ratio is also dependent on the relative
abundances of sulphur and neon, but the strong temperature dependence
should dominate in our data. We apply an extinction correction of
\av~=~14~mag to the mid-infrared data, based on the \ba/\bg\ data of
Rieke \etal\ (1980).  This large extinction only changes the
\s4/\ne2\ color by 0.26 dex (Rieke \& Lebofsky 1985) to
log(\s4/\ne2)~$=-2.7$, or a $3\sigma$ upper limit of $-1.9$.  We
compare these values to single star photoionization Cloudy (Ferland
1993) models to determine effective temperatures as a function of
ionization parameter, obtaining upper limits of 36000 K, 38000 K, and
49000 K respectively for logU~$=-1.5,\ -2.5,$ and $-3.5$.  The weak
detection of [\ion{S}{4}] on the central slit position implies
effective temperatures of 34400 K, 36000 K, and 42300 K for the
respective logU values.

The main problem with this technique is the strong dependence of the
\s4/\ne2\ ratio on ionization parameter.  The Lyman continuum flux is
$10^{53}$~s$^{-1}$ (see \S\ref{sec:uvflux}).  From
Figure~\ref{fig:nbimages}, we see that all of the \bg\ flux comes from
within a radius of 7\arcsec, or 90~pc.  From Heckman \etal\ (1990),
the nucleus has an electron density of at least 600~cm$^{-3}$ and
probably higher since dust would have hidden the nucleus at 6700~\AA.
An electron density range of 600-1000~cm$^{-3}$ gives $-$logU~$=
2.2-2.5$.  It would take a large amount of substructure in the nucleus
with a density of n$_e = 10^4$~cm$^{-3}$ to give logU~$=-3.5$.

For the logU~$=-1.5$ and logU~$=-2.5$ cases, we examined a two star
model, one star with T~=~T$_{eff}$ and the other with T~=~40000 K.  We
determined how bright the 40000 K star would have to be relative to
the T$_{eff}$ star to alter the \s4/\ne2\ ratio by 0.15 dex, or
roughly 1$\sigma$.  For the four cases: logU, T$_{eff}=-1.5$, 34400;
$-$1.5, 36000; $-$2.5, 36000; $-$2.5, 38000 we found that the 40000 K
star contributed the following percentages of the Lyman continuum
photons: 2.5; 7.4; 6.8; 31.  These percentages can be combined with
the starburst models of Section~\ref{sec:models} to determine the age
of the starburst.  As discussed below, the effective temperature is
unlikely to be as high as 38000~K, so a conservative upper limit to
the ionizing flux contribution by stars hotter than 40000~K is
probably 15\%.

Vanzi et al. (1996) suggested that the ratio HeI(1.7)/Br10 is useful
to constrain the temperature of the hot stars because it is not
heavily affected by reddening or electron temperature.  Our upper
limit for the 1.7$\mu$m HeI line (Table~\ref{tbl:helines}) corresponds
to an upper limit of HeI(1.7)/Br10 $\leq$ 0.2, placing an upper limit
of $\sim$37,000K on the stellar temperature exciting the ISM.  Carral
et al. (1994) use the far infrared fine structure lines to derive an
effective temperature for the exciting stars of $34,500 \pm 1,000$K.

Thus, three independent means of estimating the temperature of the
exciting stars from infrared spectra give consistent results of
T$_{eff} < 37,000$~K. The \oc/H$\beta$ ratio would give a value a few
thousand degrees higher than the infrared indicators. However, we show
in Section 5.4 that a significant portion of the \oc\ may be
shock-excited in supernova remnants. Also, at $\sim0.4$ solar and
higher metallicities, it is expected that \oc/H$\beta$ and other
optical indicators will overestimate the temperature of the
photoionizing field for a variety of reasons, such as an increase in
electron temperature due to depletion onto grains or to photoelectric
heating (Shields \& Kennicutt 1995). Our CLOUDY models show the
infrared-line-derived temperatures to be virtually independent of
electron temperature (c.f. Vanzi \etal\ 1996).

\subsection{Luminosity}
\label{sec:lum}

In a dusty region full of young stars (such as a nuclear starburst)
much of the stellar radiation is reprocessed by dust, so the
far-infrared luminosity serves as a reliable estimate of the
integrated output of the hot stellar population. We have estimated a
luminosity of $1.1\times10^{10}$~\lsun\ from measurements at 1.3mm
(Kr\"{u}gel \etal\ 1990), the submillimeter (Gear \etal\ 1986), the
far-infrared (Smith \& Harvey 1996, plus IRAS measurements from Rice
\etal\ 1988), and the mid-infrared (Rieke \& Low 1975). The small-beam
measurements by Smith \& Harvey proved useful in determining the
fraction of luminosity actually arising from the starburst, as the
scans of Rice \etal\ show that a non-negligible fraction of the
far-infrared luminosity in \gal\ is due to a cool extended component.
This value should be considered a lower limit to the total luminosity,
since a significant amount of energy may be escaping perpendicular to
the plane of the galaxy.

We have computed a lower limit to the K-band luminosity of the
underlying stellar component by fitting an exponential disk model to
the K-band light at large radii (c.f. Forbes \& Depoy 1992) and
computing the integrated light contribution of this model in the
central 15\arcsec.  The absolute K magnitude of this component is
$-$18.3, less than 15\% of the total K-band light after accounting for
extinction.  Virtually any more sophisticated model, such as allowing
the old stars to be more centrally concentrated, will increase this
proportion.

Using BC$_{\rm K} = 2.7$ and M$_{\rm Bol}(\odot) = 4.75$, we compute
that the bolometric luminosity of the underlying population is less
than $2\times10^8$~\lsun, or less than 2\% of the observed value, so
we will ignore this small contribution to the total. However, we have
subtracted the K flux from the older stars from the total to derive
the expected starburst output at this wavelength, and it is this
corrected value which appears in Table~\ref{tbl:summ}.

\subsection{Mass}
\label{sec:mass}

\subsubsection{Rotation}
An abundance of kinematic data exists for the nuclear region of \gal.
Several rotation curves exist in the literature that have been
obtained at radio wavelengths, in molecular emission lines such as CO
(e.g., Mauersberger \etal\ 1996, Canzian \etal\ 1988) or CS (Peng
\etal\ 1996) or in hydrogen recombination lines (Anantharamaiah \&
Goss 1996).  While most of these observations have high spectral
resolution, they typically have poor spatial resolution and are not
always useful for probing the dynamics very close to the nucleus.  The
data of Anantharamaiah \& Goss are an exception --- their data
have a spatial scale similar to ours and indicate a mass of
$3\times10^8$~\msun\ inside a radius of 5\arcsec, comparable to the
mass we derive in the next section.

Optical rotation curves of the nuclear region also exist (e.g.,
Mu\~{n}oz-Tu\~{n}on, Vilchez, \& Casta\~{n}eda 1993 and Arnaboldi
\etal\ 1995).  The observations by Mu\~{n}oz-Tu\~{n}on \etal\ are likely
affected by dust extinction in the sense that they indicate a rotation
curve that is too shallow (as demonstrated by Prada \etal\ 1996).  
The measurements by Arnaboldi \etal\ were obtained at higher
spatial resolution and indicate a slope for the rotation curve that
is somewhat higher than indicated by our data (see below).

In Figure~\ref{fig:rotcurve} we present rotation curves in strong
emission lines and in the (2,0) rovibrational band of CO at
2.3\micron.  The stellar CO feature has the advantage that it traces a
component of the galaxy nucleus which is not affected by
non-gravitational processes such as shocks or winds.  The stellar
feature also traces a component of the population which is smoothly
distributed throughout the region of interest, as indicated by our
K-band image. Tracers such as \bg\ and \h2\ are likely concentrated in
discrete HII regions and the surfaces of molecular clouds, although
the rotation curves derived from both the gaseous and stellar features
match quite well in the central region.

The velocity gradients across the nucleus along the major axis are
tabulated for our three chosen near-infrared features in
Table~\ref{tbl:velgrad}.  They overlap within the uncertainties, and
the combined average is $7.5\pm0.3$~\kms/\arcsec.  This number is not
corrected for inclination.  Our infrared emission lines probably trace
a compact nuclear disk of gas. Although there is no way to be certain
that the inclination of the nuclear region follows that of the rest of
the galaxy, the only correction we can reasonably make is to correct
for this inclination, which is 78\arcdeg\ (Pence 1981). This
correction gives us a value of $7.7\pm0.3$~\kms/\arcsec.  This value
is similar to that from Puxley \& Brand (1995) but somewhat lower than
we find from the data of Prada \etal\ (1996), from which we calculate
$7.8\pm0.4$ and $9.8\pm0.2$, respectively, if we eliminate points that
are obviously not on a linearly-rising rotation curve.  The rotation
curve indicated by all the near-infrared measurements is quite
shallow, however, and indicates a mass of $1.7\times10^8$~\msun\
within a 10\arcsec\ radius (the radius at which the rotation curve
turns over) if we assume the mass is distributed spherically.

We have also obtained a rotation curve for the well-studied galaxy
NGC~3115 as an evaluation of our techniques to measure galaxy rotation. 
In Figure~\ref{fig:rotcurve}, we compare our results with the
optical data of Kormendy \& Richstone (1992).  The rotation curves
compare quite well, although the Kormendy \& Richstone rotation curve
rises more steeply in the center, most likely due to better seeing and
their smaller pixel size (0\farcs435) as compared with our 1\farcs8
pixels at the 1.55m telescope.  This comparison gives us confidence
that we are correctly measuring the galaxy dynamics, despite
the fact that the CO band is a broad and one-sided feature.

\subsubsection{Dispersion}
\label{sec:disp}

A more robust way to determine the mass is to measure the velocity
dispersion of the stars and compute the mass by assuming a model for
the stellar distribution, as in Shier \etal\ (1994).  Using the
cross-correlation routines described by Shier \etal\ , we measure a
stellar velocity dispersion in the CO(2,0) feature of $87\pm10$~\kms\
in the 12\arcsec\ aperture we used to extract the spectra, where the
error bars include both the statistical uncertainty as measured by a
Monte Carlo simulation and an estimate of the systematic uncertainty
as determined by experiments with fitting standard stars.  The
template star for this calculation was HR~7475.

The observed velocity dispersion is the sum of the dispersions in each
extracted pixel, weighted by the luminosity profile of the galaxy.  We
fit a profile to the 2.3\micron\ region of our spectrum of \gal,
modeling the light distribution using the $\eta$ models of Tremaine
\etal\ (1994).  The model that best fit the luminosity profile has a
scale radius of 6\arcsec\ (about 73~pc at the distance of \gal) and
$\eta=3$.  Using this galaxy profile, we measure a mass of $3.9 \pm
0.9 \times10^8$~\msun\ within a 7\farcs5 radius.  The uncertainties in
the mass determination are dominated by uncertainties in the velocity
dispersion, as the derived mass depends on the square of the
velocity.

\subsubsection{Mass Budget}
\label{sec:mass-budget}

To determine the mass we may allocate to star formation in a recent
starburst, we must subtract from the dynamical mass the mass of the
old stars and molecular gas in the nucleus.  Assuming the same
conversion from mass to K-band luminosity determined by ERRL96 and the
K-band luminosity derived in \S\ref{sec:lum}, this preexisting stellar
population is expected to have a mass of at least
$3.3\times10^8$~\msun, and is likely to be large enough that it
accounts for virtually all the stellar mass in the nucleus.  The large
proportion of mass in old stars confirms the theoretical predictions
that a starburst will be triggered when only 10-20\% of the nuclear
mass is in the form of gas (Wada \& Habe 1992; Bekki 1995).  Using the
standard CO to gas mass conversion, the molecular gas mass may be as
large as $1.4\times10^8$~\msun\ (Mauersberger \etal\ 1996).
Uncertainties in the application of this conversion to starburst
galaxies (e.g., Maloney \& Black 1988; Shier \etal\ 1994) led
Mauersberger \etal\ to propose a much smaller mass of $\sim 3 \times
10^7$~\msun\, based on measurements of dust continuum and a rare
isotope of CO.

The lower limit on the mass in old stars and the gas combined leave no
more than $0.3 \pm 0.9 \times 10^8$~\msun, or a strong upper limit of
$2.1\times10^8$~\msun\ on the starburst mass; it is likely much less.
We show in \S\ref{sec:models} that the best-fitting starburst models
indicate a mass of a few times $10^7$~\msun.

\subsection{Ionizing Flux}
\label{sec:uvflux}

To estimate Q(H) (number of hydrogen ionizations per second), we correct the
longest-wavelength (i.e., the line least affected by extinction)
hydrogen recombination line flux we observe (\bg) for extinction and
convert it to Q(H) using the ratio $I({\rm H}\beta)/I({\rm
Br}\gamma)=30.3$ (from Hummer \& Storey 1987) and $\alpha_{\rm
B}/\alpha^{eff}_{\rm H \beta}=8.40.$ From our narrow-band 
images, we determine the \bg\ flux in a
15\arcsec\ aperture to be $9.16\times10^{-13}$~\cgs.  We correct this
flux to $2.14\times10^{-12}$~\cgs\ using $A_{K}=0.9$ from
\S\ref{sec:extinct}.  The value of Q(H) implied by this \bg\ flux is
$1.0\times10^{53}$~s$^{-1}$.

We obtain a similar value from our \ne2\ measurement.  The dereddened
(\av=14 mag) \ne2\ flux is (6.6$\pm$0.2) $\times$ 10$^{-11}$ \cgs.
Combined with the formula of Roche \etal\ (1991) and a distance of
2.5~Mpc, this gives \nlyc~$=1.3 \times 10^{53}$~photons~s$^{-1}$.  The
Roche \etal\ formula assumes a solar abundance of neon.  While the
actual neon abundance in \gal\ is unknown, the fact that this number
is within 30\% of the value derived from the \bg\ flux suggests that
the neon abundance is not too different from solar.

We obtain a very similar Q(H) from the thermal radio flux measured by
Turner \& Ho (1983), using their equation for converting the thermal
component of the 6cm radio flux to Q(H).  They estimate that 125mJy,
or less than 10\% of the total flux at 6cm, is due to thermal
emission, implying Q(H)~$\sim8\times10^{52}$~s$^{-1}$.  A somewhat
higher estimate, Q(H)~$=3.7\pm0.8 \times 10^{53}$~s$^{-1}$, is derived
by Puxley et al. (1997) from mm-wave H recombination lines.
 
These numbers would be approximately equal to the total number of
ionizing photons per second, \nlyc, were it not for the presence of
significant amounts of dust in the starburst region which can compete
with hydrogen atoms in absorbing ionizing photons. As such, we can
take Q(H) to be a lower limit on \nlyc, i.e., log (\nlyc)~$\gtrsim53$.

\subsection{CO index}

We have listed the equivalent widths of the CO lines, along with other
absorption features, in Table~\ref{tbl:abs}.  Using the definition of
the CO index by Kleinmann \& Hall (1986), where we have redshifted the
wavelengths by 245~\kms, our spectroscopic CO index is 0.31, which
corresponds to a photometric index of 0.18.  If we deredden the
spectra using the extinction described in \S\ref{sec:extinct}, our
spectroscopic index is 0.32, for a photometric index of 0.19. Using
the specification of Doyon, Joseph, \& Wright (1994a), we measure
$CO_{sp}=0.29$, which gives $CO_{ph}=0.21$.

Although both approaches agree well, the Doyon \etal\ approach is more
appropriate for data of lower resolution than ours: we will therefore
use the Kleinmann and Hall definition.  We will correct our measured
value for dilution by an underlying population by assuming that the
old stellar population contributes $\sim 15$\% of the K-band light
(see \S\ref{sec:lum}) and that the old population has a CO index of
0.15 (Frogel \etal\ 1978), giving us an intrinsic CO index of 0.20 for
the starburst population.

\subsection{Other Constraints}

Fabbiano (1986) shows that \gal\ has a large X-ray feature out of the
plane of the galaxy and extending at least to 7 kpc.  She suggests
that this source is due to hot gas escaping from the galaxy in a
starburst-driven superwind. A similar feature in M~82 was used in
RLRT93 to place timing constraints on the starburst in that galaxy,
based on the interval from the initiation of supernova explosions to
when the hot gas could expand to create the feature. Similar
constraints should hold for \gal. For example, in a simple
single-burst model, an interval of about 8 Myr must pass from the peak
of the star formation to allow gas to expand out of the starburst
region in \gal.

As in M~82, it appears that the metallicity in the \gal\ starburst is
approximately solar or less than solar rather than highly enriched
(e.g., Carral et al. 1994; Ptak et al. 1997). The oxygen abundance is
of particular interest because this element is released copiously in
supernovae (Arnett 1978). Although RLRT93 speculated that highly
ionized oxygen might hide in the hot superwind, recent X-ray spectra
indicate that, if anything, the abundance is less than solar in this
plasma (Ptak et al. 1997). Although there is some controversy about
the low metallicities derived from X-ray data (Fabbiano 1996), any
model for the starburst that produces high oxygen abundances is likely
to be inconsistent with observation.

 \section{Starburst models}
\label{sec:models}

\subsection{Description of Models}
RLRT93 fitted the starburst in M~82 using the models of Rieke \etal\
(1980) with updated stellar tracks and atmospheric parameters.  The
starburst model uses the grid of stellar evolution tracks of Maeder
(1992), which have been assigned observational parameters based on a
combination of atmosphere models and empirical calibration.  The
models include stars up to 80\msun\ and interpolate between the tracks to reduce
oscillations caused by discreteness in the stellar masses. 
Further details can be found in RLRT93.

\subsection{\gal\ models}
\label{sec:n253-models}

\begin{figure}
\plotone{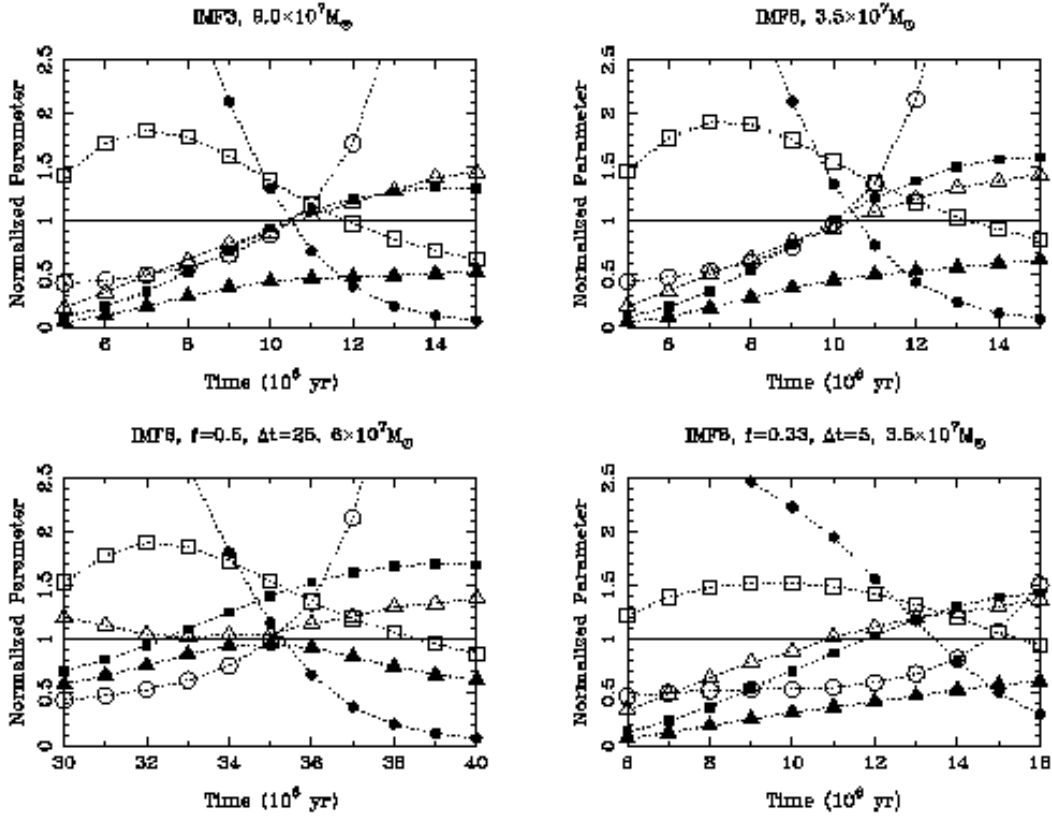}
\caption{Starburst models of \gal\ using a solar-neighborhood IMF
(RLRT93's IMF~3) and an IMF biased towards massive-star formation
(RLRT93's IMF~8).  The points represent output from our starburst
model as a function of time, where each curve has been ratioed to the
corresponding observational parameter listed in Table~\ref{tbl:summ}.
A reasonable fit is indicated where all the curves come within the
range of uncertainties to match the target value (ideally at 1).  The
IMF used and the mass required to match the observations is indicated
for each panel.  For the bottom panels, double-burst models with a
delay $\Delta$t (in millions of years) between bursts and the fraction
f of the mass that was contained in the second burst are presented.
The symbols represent the following quantities: open square = \lbol,
open triangle = CO index, open circle = T(UV)$_{40}$, filled triangle
= SN rate, filled square = L$_{\rm K}$, filled circle = \nlyc.}
\label{fig:models}
\end{figure}

For \gal, the model values of \nlyc, \lbol, CO index, K-band flux and
$\nu_{\rm SN}$ (supernova rate) are displayed as a function of time in
Figure~\ref{fig:models}.  Each point along the curves has been divided
by the observed values for \gal\ in Table~\ref{tbl:summ}, so the
target value for each quantity is 1.  A fit can be considered good
when all curves meet the target value simultaneously.  In practice,
since some of the observational parameters are uncertain or are merely
lower limits, we choose the point on the plot where the curves can
simultaneously meet the target values within the specified range of
uncertainty.

The various input parameters constrain the starburst model in
different ways, depending on how each is determined.  Our value for \nlyc\ 
must be considered a lower limit. Given the uncertainties due to poor
angular resolution in the far infrared and to energy escaping perpendicular
to the plane of the galaxy, we require our models to match the
observed \lbol\ to within 50\%.  The formal uncertainties in our
determination of the CO index and K-band flux are small, but due to
uncertainties about continuum placement and corrections for extinction,
we assign uncertainties to these quantities
of 20\%.  The supernova rate is probably no more certain than a factor of two;
however, since our estimate is generally on the low side 
compared with those derived by other
means, and this discrepancy cannot be removed in the starburst models
by changing the distance to the galaxy (since the supernova rate and
other luminosity-derived parameters will scale together), we prefer
models that do not fall significantly below the estimated rate. We have 
also made use of the temperature constraint derived in
\S\ref{sec:temp}.  In the figure, we plot a curve labeled
T(UV)$_{40}$, which is the ratio of the total ionizing flux to the
ionizing flux produced by stars hotter than 40,000~K.  The starburst
model is required to age sufficiently that no more than 15\% 
of the ionizing flux comes from stars hotter than 40,000~K.

We created two models with a simple, single-burst
model, with the star formation rate a Gaussian in time with a FWHM of
5 million years.  The models begin 5 million years before the peak of
the star formation.  This short Gaussian burst of star formation is
very efficient in converting stellar mass to luminous output. 

Our first model used a solar-neighborhood IMF (RLRT93's IMF3).  Using
a mass of $9 \times 10^7$~\msun, the model comes closest to the
constraints at a time of 5.5 million years after the peak of the star
formation.  The CO index, the ionizing flux, and K-band luminosity are
very close to the observed values, while the bolometric luminosity is
a factor of 30\% higher than observed and the supernova rate is a
factor of 2 too low.  There has been inadequate time for supernovae to
eject a hot plasma to the observed extent of the X-ray emission. This
model falls short of the observational constraints, despite our
conservatism in setting them. The model also makes the uncomfortable
prediction that the star formation efficiency is 40-75\%, assuming the
mass not used to make stars is still in the form of molecular gas.

We also attempted to model the starburst using an IMF more biased towards
the formation of massive stars.  Such a model produces much more
luminosity for a given mass and should do a better job of fitting the
observations of \gal\ while remaining within the mass constraint.  We
used the IMF found by RLRT93 to fit the M~82 observations best (their
IMF8).  This model looks very similar to the IMF3 one, but it only uses
$3.5\times10^7$~\msun.

\begin{figure}
\plotone{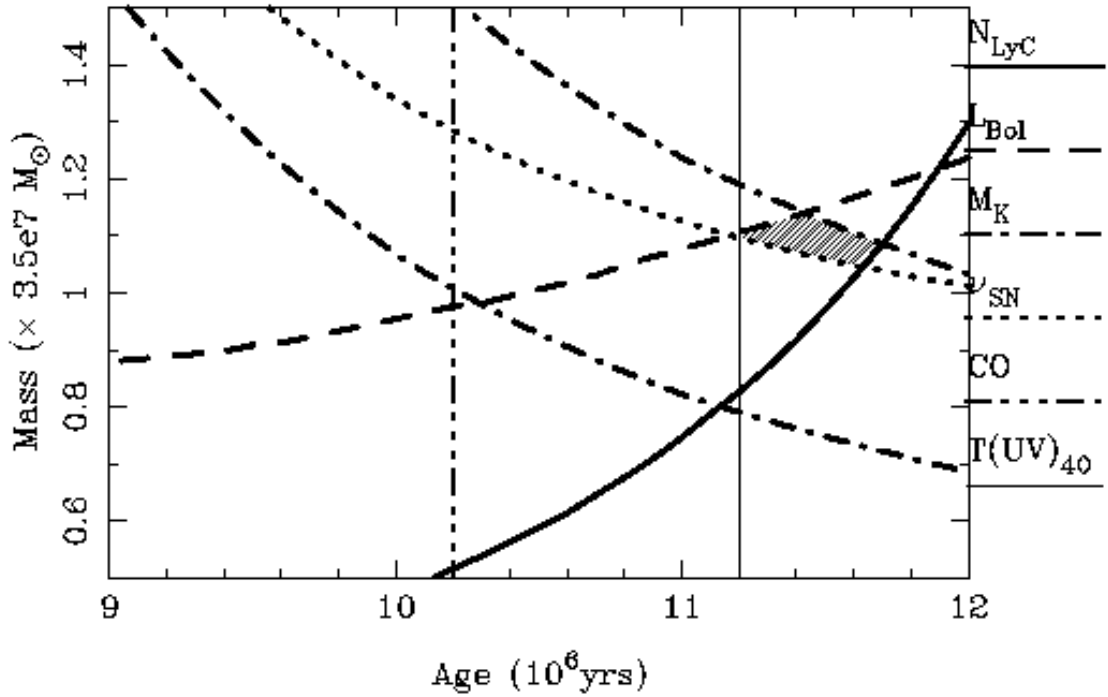}
\caption{Allowed values in the (mass, age) plane for a single-burst
starburst model using IMF8 of RLTLT93.  The hatched region is
constrained by the model values from Figure~\ref{fig:models} with
uncertainty ranges as described in the text.  The CO index and
T(UV)$_{40}$ parameters do not scale with mass and therefore serve
primarily as age constraints.}
\label{fig:unc}
\end{figure}

To illustrate how the uncertainties in the various parameters affect
the determination of the mass and age of the starburst, in
Figure~\ref{fig:unc} we have plotted the range of allowed values for
each quantity constraining the starburst for the IMF8 model.  For
example, the two dot-dashed lines in the plot indicate a range of 20\%
around the nominal K-band luminosity.  Quantities such as the CO index
and T(UV)$_{40}$ do not scale with mass and so serve primarily as age
constraints, showing up as vertical lines on this plot, where the
T(UV)$_{40}$ is a lower limit and the CO index has an allowed range
which extends off the far end of the plot.  The hatched region
indicates the space in the (mass, age) plane that can be occupied by
the model while still matching the observations.  In this case, the
primary constraints are the lower limits on the supernova rate and the
UV flux and the upper limits on the K-band luminosity and the
bolometric luminosity.  The model falls a little short of meeting the
constraints at $3.5\times10^7$~\msun---a better fit occurs at 6
million years after the peak of star formation, using a mass of
$3.8\times10^7$~\msun.  However, the figure illustrates that no single
burst model approaches the X-ray timing constraint (at $\sim 13$~Myr,
8~Myr after the peak of star formation).

We next explored the effect of varying the star formation history by
adding a second burst.  We have parameterized this second burst by a
time delay ($\Delta$t) after the initial burst and a fraction (f) of
the mass consumed by both bursts that goes into the second burst, so
that f~=~0.5 implies identical masses for both bursts. Although the
additional parameters introduced by a two-burst model can help fit
many constraints, they invariably require increased mass
(RLRT93). Consequently, since the IMF3 model already used a lot of
mass, we have not computed double burst models for it. In
Figure~\ref{fig:models} we display a model with two IMF8 bursts of
equal strength, separated by 25 million years.  Since we have diluted
the mass available by putting it into two widely-separated bursts,
this model requires a total mass of $6\times10^7$~\msun.  The model
achieves a good fit at a time of 34.5 million years.  The two main
advantages of this double-burst model are that there is adequate time
for expansion of the X-ray superwind and that the supernova rate is in
reasonable agreement with the target value.  Models with shorter time
intervals between bursts predict supernova rates below the target
value.

Although for convenience in modeling, we have used two separated
bursts in this model, similar results would be obtained with other
time dependencies for the star formation rate, so long as: 1.) a
substantial portion of stars formed about 30 million years ago, so
they can produce supernovae without ionizing flux; and 2.) the star
formation rate has decayed rapidly for the last roughly 5 million
years, so the correct ionizing flux can be produced without an excess
of very hot stars.

\subsection{Residual Problems}

These three models only meet the temperature constraint with
difficulty - if we reduce the allowed fraction of ionizing photons
emitted by stars hotter than 40000~K from 15\% to 10\%, significantly
more mass is required.  The temperature constraint is satisfied
trivially if we arbitrarily impose an upper-mass cutoff on the models.
This explanation is unlikely, however, since observations of very
young star formation regions always indicate that very massive stars
are formed (e.g., Conti, Leitherer, \& Vacca 1996; Leitherer
\etal\ 1996).  Extending the duration of the burst of star
formation, as in the model we present in the last panel, only
aggravates the temperature problem, as a larger fraction of the
ionizing flux is produced by recently-formed massive stars. It might
be thought that the dusty environments of starbursts would absorb
hard UV photons and soften the ionizing spectrum. However, the
absorption properties of dust are believed to {\it harden} the UV
ionizing spectra (Aannestad 1989; Shields \& Kennicutt 1995).

The lack of extreme oxygen over abundance is a strong constraint on
the shape of starburst IMFs at the highest masses (RLTRT93).  The
evidence from X-ray spectra that the oxygen may be underabundant
relative to solar in \gal\ (and M~82) (Ptak \etal\ 1997) therefore may
add a difficult constraint to the models, unless the amount of oxygen
returned to the ISM is lower than expected.

\subsection{The ``Biased'' Initial Mass Function in \gal}

Although the case is not as definitive as in extreme galaxies like
M~82 (RLRT93) and NGC 1614 (Shier, Rieke, \& Rieke 1996), \gal\
appears to be another example of a starburst that cannot be modelled
easily with a "local" IMF such as that of Scalo (1986).  IMF8 was
adopted for M~82 after consideration of a broad range of alternatives,
and appears to be a satisfactory solution to this problem for \gal\
also.

The presence of a starburst IMF biased toward high mass stars
continues to be controversial (e.g., Satyapal \etal\ 1997). However,
many studies have fitted starburst observations with a "Salpeter" IMF,
with a slope of -1.35 but with a turnover at low masses to satisfy
observations of the local stellar population (e.g., Conti, Leitherer,
\& Vacca 1996; Satyapal \etal\ 1997). This IMF is virtually identical
in observable properties to IMF8 and differs just as strongly from the
IMF of Scalo (1986)!  For example, if the "Salpeter" IMF adopted by
Satyapal et al.  (1997) is normalized to IMF8 through the requirement
that the stellar populations have equal total mass, the two IMFs are
the same in the critical 15-30 M$_\odot$ region except for a small
difference in slope. The "Salpeter" IMF predicts a larger number of
very massive stars than does IMF8, and would exacerbate the problem
with oxygen abundance.

There are substantial difficulties in determining the local IMF, and
the IMF above $\sim 1$~\msun\ in other environments shows substantial
variations either from intrinsic differences or interpretive errors
(Scalo 1998). The debate about biased IMFs in starbursts is therefore
no longer very relevant. The most interesting development is how all
investigators of starburst galaxies have converged on an IMF very
similar to IMF8 of RLRT93.

\section{Other results}
\label{sec:other}

\subsection{Continuum morphology}

A striking feature of the images in Figure~\ref{fig:bbimages} 
is the bright, elongated feature centered on the
nucleus and oriented roughly east-west.  This feature was first noted by
Scoville \etal\ (1985), who describe it as a bar; Pompea \& Rieke
(1990) found that it curves as it approaches the nucleus and suggested
it might instead be inner spiral arms.  Our deep infrared images
indicate that this apparent curvature arises from a
zone of strong extinction.  Our dereddened K-band image presented in
Figure~\ref{fig:deproj} leaves little doubt it is a bar.

Also evident is a ring with a diameter of $\sim350$\arcsec\
($\sim4$~kpc) along the major axis of the galaxy.  This ring is very
nearly circular, as can be seen in our deprojected image of the galaxy
in Figure~\ref{fig:deproj}.  The ring is also apparent in the deep
H-band image from Forbes \& DePoy (1992), although they do not discuss
it. The bar crosses the ring and also has a length of
$\sim350$\arcsec.

Canzian, Mundy, \& Scoville (1988) show a bar in CO emission that is
about 30\arcsec\ long and oriented along the stellar bar discussed
above. Thus, the configuration of molecular gas in the nucleus of
\gal\ approximates the theoretical predictions that mutual torques
between stellar and gas bars will allow the gas to lose angular
momentum and sink into the nucleus, where it concentrates until
instabilities set in that trigger a starburst.

\subsection{Emission-line morphology}

In Figure~\ref{fig:pv} we present position-velocity diagrams of some
strong emission lines in \gal. These images and the the narrow-band
images in Figure~\ref{fig:nbimages} show that the starburst is not
distributed symmetrically along the major axis but instead is
concentrated in a region north-east of the nucleus.  They also show
significant differences from line to line. The H recombination and
\ne2\ have a very similar asymmetric distribution in position/velocity
space.  The \fe2\ is distributed symmetrically to both sides of the
nucleus.  The \h2\ emission is distributed over a broader region still
($\sim 500$~pc). These behaviors are consistent with the starburst
originally being fairly uniformly distributed around the nucleus, as
indicated by the supernovae traced by \fe2, but now being active
primarily on the nucleus and to the northeast. They also indicate that
the \h2 emission is partially excited by a mechanism that extends
outside the other starburst indicators, e.g., molecular cloud
collisions.

\subsection{\h2\ emission}
\label{sec:h2}

The excitation mechanism of \h2\ in starburst galaxies is under
debate.  The strong UV flux should excite a significant 
amount of \h2.  However, many of these
galaxies display strong emission in the (1,0)S(1) line (e.g., Goldader
\etal\ 1996) but not in other \h2\ lines, suggesting that collisional
processes rather than fluorescent ones dominate (c.f., Black \& van
Dishoeck 1987).

One reason that UV-excited \h2\ is so difficult to detect is that its
energy is emitted in a huge number of lines, whereas thermally excited
\h2\ emits most of its luminosity in just a few.  In low density
gas, fluorescent \h2\ emits just 1.6\% of its infrared luminosity in
the readily observed (1,0)S(1) line while thermally excited \h2\ emits
9\% of its infrared luminosity in this line.  Despite the difficulties
associated with detecting a number of very weak lines, fluorescent
\h2\ has been detected recently in some galaxies (Doyon
\etal\ 1994; Kulesa \etal\ 1998.)

The combination of high resolution and sensitivity, broad spectral
coverage, and our technique for subtracting the stellar continuum
allow us to measure an unprecedented number of \h2\ lines in an
extragalactic source (Table~\ref{tbl:h2lines}).  We have used the
A-values, energy levels, and statistical weights adopted by Ramsay
\etal\ (1993) to derive the temperatures listed in
Table~\ref{tbl:rottemp}.  The temperature derived from the two
strongest lines, (1,0)S(1) and (1,0)S(3), is consistent with 2000~K
and not with 1000~K. In Table~\ref{tbl:h2lines}, we have also listed a
set of line ratios dereddened using the extinction derived in
\S\ref{sec:extinct}.  We compare these line ratios to two \h2\ models
from Black \& van Dishoeck (1987)---a UV-excited model (their model
14), plus a collisionally-excited model appropriate for 2000~K (their
model S2).

A first glance would suggest that the excitation of the \h2\ is
dominated by collisions, rather than UV pumping. Following the
arguments of Kulesa \etal\ (1998), however, we can show that
fluorescent emission plays an important role in producing the \h2\
spectrum of \gal.  In the table we present a mixed model in which 25\%
of the flux in the (1,0)S(1) line is due to fluorescence, modifying
the other line ratios accordingly.  One can see that this model fits
the data significantly better, without looking radically different
from a pure thermal model. The ability to distinguish between pure
thermal and mixed thermal and fluorescent
\h2\ emission depends on having high-enough sensitivity and resolution
to detect faint lines.  

For a quantitative comparison, we measure a flux in the (1,0)S(1) line
of $1.2\times10^{-13}$~\cgs\ in our 2\farcs4~$\times$12\arcsec\
spectroscopic aperture.  We correct this value to a 15\arcsec\
reference aperture using the narrow-band \h2\ image and obtain a flux
of $3.5\times10^{-13}$~\cgs.  Correcting this value for extinction
gives us $8.4\times10^{-13}$~\cgs, or $1.6\times10^5$~\lsun\ at a
distance of 2.5~Mpc.  Using this luminosity, the total \h2\ luminosity
in the mixed model is $3.8\times10^6$~\lsun---$1.3\times10^6$~\lsun\
from thermal emission and $2.5\times10^6$~\lsun\ due to fluorescent
emission; that is, nearly two thirds of the total \h2\ infrared
luminosity is derived from fluorescence.

Luhman \etal\ (1994) report global observations of \h2\ in Orion, with
the result that about 1900~\lsun\ are due to fluorescent emission
while only 40~\lsun\ are due to thermal emission.  From these numbers,
we calculate that $1.3\times10^3$ HII regions similar to Orion would
be required to account for the fluorescent emission we observe in
\gal.  This number of Orion-like HII regions is also nearly enough to
account for the ionizing flux from \gal---$3 \times 10^{49}$~s$^{-1}$
per Orion HII region (Bertoldi \& Draine 1996) yields half the \nlyc\
for the starburst.

These HII regions would provide $5.3\times10^4$~\lsun\ of thermal
emission, leaving nearly all the thermal emission from \gal\
unaccounted for.  The deficit could be made up by thermal emission
from diffuse shocks, either associated with supernova remnants or on
larger scales.  Graham, Wright, \& Longmore (1987) and Burton \etal\
(1988) derive an \h2\ luminosity in the supernova remnant IC~443 of
$\sim70$~\lsun.  Taking this to be the average luminosity per
supernova remnant, and given the estimated supernova rate in \gal, the
average \h2-emitting lifetime of the supernova remnants would have to
be $\sim$ $6\times10^5$ years. This lifetime is implausibly long,
particularly since at this age any supernova remnant in \gal\ will
have expanded and merged with the general hot interstellar
medium. Instead, we suggest that much of the thermal \h2\ arises from
cloud-cloud collisions or other diffuse shocks (c.f. van der Werf
\etal\ 1993), some of which may be powered by mechanical energy
released by supernovae and stellar winds. This possibility is
supported by the differences in distribution of \fe2\ and \h2\
emission in the position-velocity plots in Figure~\ref{fig:pv}.

\subsection{Nature of Weak-\oa\ LINERs}
\label{sec:liners}

Although the intrinsic \ob/\oc 5007 line ratio is difficult to
determine for \gal\ because of the strong reddening, other, less
reddening-dependent diagnostic optical line ratios indicate it is a
transitional HII/weak-\oa\ LINER. Filippenko \& Terlevich (1992) use
nebular models to demonstrate that this class of galaxy can be excited
by hot stars, with T~$\ge45,000$~K. The locus of \gal\ on the
diagnostic plots of Filippenko \& Terlevich falls even closer to their
theoretical calculations than does the zone they indicate for typical
weak-\oa\ LINERs. Alternatively, it has been proposed that these LINERs
are excited by shocks (e.g., Heckman 1980).

The strong arguments against the effective stellar temperature in
\gal\ being nearly as hot as 45,000~K are difficult to reconcile
with the model of Filippenko \& Terlevich. In fact, our data are
difficult to reconcile with any model based on photoionization
from a hard UV spectrum, since the indications of low stellar
temperature also argue against a hard UV field from a mini-AGN or
other source. Moreover, there are few other traces of a mini-AGN
in this galaxy. For example, our measurement of Br$\gamma$ shows
the line to be narrow with no broad wings. On the extinction-free
\oa/H$\alpha$ vs. \fe2\ 1.64$\mu$m/Br$\gamma$ diagnostic
diagram, \gal\ falls near the middle of the starburst locus
(Alonso-Herrero et al. 1997). There is no hard X-ray source in
the nucleus of the galaxy, and although there is a compact radio
source (Turner \& Ho 1985), it is of low luminosity. 

It seems likely that the LINER-like characteristics of \gal\ are produced
by a mechanism other than photoionization. We have shown that the
\fe2\ emission is likely to be produced directly in supernova remnants.
Alonso-Herrero et al. (1997) demonstrate that \gal\ falls
approximately on the mixing line between HII regions and supernova
remnants on the \oa/H$\alpha$ vs \fe2/\bg\ diagram. That is, the \oa\
strength is also consistent with an origin predominantly in the same
supernova remnants that produce the \fe2, with additional H
recombination emission from HII regions.

The influence of supernovae on the observed emission line spectrum of
\gal\ can be determined more quantitatively. We have determined an
average supernova emission line spectrum by averaging the line
strengths (normalized to H$\alpha$) for 32 supernova remnants measured
by Danziger \& Leibowitz (1985), Fesen, Blair \& Kirshner (1985), and
Blair \& Kirshner (1985). Similarly, an average ratio of \fe2\
1.64$\mu$m/\bg\ can be determined for supernovae measured by Oliva,
Moorwood \& Danziger (1989; 1990).  Taking H$\alpha$/\bg~ $=103$ from
Case B, we find that \oa/\fe2~$\sim 1.25$. Relating these values
through the H recombination lines is an essential step because of the
differing geometries in the available optical and near infrared
spectra.  Nonetheless, because of the relatively small amount of
infrared data, the derived ratio is somewhat uncertain.

We can predict the optical spectrum from the infrared one with: 1.)
the relations derived in the preceding paragraph; 2.) the total \fe2\
flux of $3.8\times10^{-12}$~\cgs\ derived in Section 3.2; 3.) the
total \bg\ flux of $2.14\times10^{-12}$~\cgs\ derived in Section 5.4;
4.) case B ratios for the H recombination lines; and 5.) standard
nebular calculations for the relative strengths of photoionized lines
in the optical (we have used the work of Shields \& Kennicutt (1995),
which includes the effects of dust in metal-rich HII regions). We
assume a stellar temperature of 38,000~K (the lowest for which Shields
\& Kennicutt give calculations) and solar metallicity. We also use the
\oa/H$\alpha$ ratio for the Orion nebula to represent a typical HII
region. The results of these predictions are summarized in
Table~\ref{tbl:nebular}.

The agreement between predicted and observed line ratios is well
within the uncertainties. About 85\% of the \oa\ and half of the
\oc\ are excited in the supernovae remnants, whereas most of the
[\ion{N}{2}], [\ion{S}{2}], and H recombination are excited by
photoionization. Although the photoionized portion of the \oc\ will
decrease rapidly with decreasing stellar temperature, the large
portion of shock-excited \oc\ will maintain good agreement with the
observations. If the intrinsic \oa/\fe2\ ratio is a factor of 1.5
higher than we have used (which is within our guess of the errors),
the agreement between predictions and observations is improved and the
\oc/H$\beta$ ratio becomes almost independent of a decreased
stellar temperature.

Thus, a combination of a metal-rich HII region photoionization model
with the shocked optical line strengths predicted by assuming the
\fe2\ is produced by supernovae provides a satisfactory fit to all the
LINER-like characteristics of  \gal.  This fit involves no free
parameters, other than those determined independently of the LINER
characteristics. The Filippenko \& Terlevich (1992) hot star
photoionization model would predict that LINERs appear as a very early
stage in a starburst. Instead, we suggest that the LINER
characteristics emerge after the stars emitting most vigorously in the
UV have died, so the HII region characteristics will fade sufficiently
to reveal the supernova shock excitation.

The decay time for ionizing flux in a starburst is only a few million
years, whereas the supernova rate is maintained for about 30 million
years. Therefore, if the star-forming episodes in starbursts are
typically short in duration, of order 20 million years, then there
should be a significant population of objects in the transitional
stage that produces LINER characteristics. In agreement with this
conclusion, we found in \S\ref{sec:n253-models} that starburst models
with these characteristics give the best overall fit to the properties
of \gal.

\section{Conclusion}
\label{sec:conclude}

We have presented high-quality spectra of the nuclear region of \gal\
in the J, H, and K bands. We have constructed a composite stellar
spectrum and subtracted it from that of \gal\ to obtain accurate
measurements of faint emission lines free from the spectral structure
due to stellar absorptions. We have also presented J, H, and
K$_s$-band images, narrow-band images in the lines of \bg\ and
\h2(1,0)S(1), spectra at 10.5\micron, and a high-resolution spectral
map at 12.8\micron.  We have used these data and data from the
literature to perform a detailed study of the interstellar medium and
nuclear starburst in \gal.  All of the observed properties, from the
radio through the X-ray, can be explained plausibly by a starburst.
Specifically, we find:

\noindent 1.) The density in the \fe2-emitting region is $\sim 5
\times 10^3$ cm$^{-3}$, similar to that observed for \fe2-emitting
regions in supernova remnants. The strength of the \fe2\ emission is
consistent with the supernova rate estimated from radio observations
and predicted by starburst models.  Together, these results confirm
that the \fe2\ is produced predominantly in the supernovae in the
starburst.

\noindent 2.) Roughly two thirds of the infrared
\h2\ luminosity is due to UV fluorescence. It is likely that
significant fluorescent-excited components of \h2\ emission in other
starburst galaxies have been missed because the distribution of
emission over many relatively faint lines makes detection difficult in
spectra of modest resolution and signal to noise.  If the Lyman
continuum is produced by Orion-like HII regions, they would also come
within a factor of $\sim 2$ of producing the fluorescent \h2.

\noindent 3.) From the MIR fine-structure lines of [\ion{S}{4}] and
\ne2, we find that the stars photoionizing the gas in \gal\
have T$_{eff} < 37000$~K, with $<15$\% of the ionizing flux coming
from stars with T~$>40000$~K.  We also use an upper limit to
HeI(1.7)/Br10 to place an upper limit on the stellar temperature of
$\sim37,000$~K.  These values are consistent with the temperature
determination of $34,500 \pm 1,000$K derived from far infrared fine
structure lines by Carral et al. (1994).

\noindent 4.) The optical line ratios in \gal\ indicate it is a
HII/weak-\oa\ LINER and they are fitted well by the hot-star ($T \sim
45,000$K) photoionization models proposed for this type of active
galaxy by Filippenko \& Terlevich (1992).  However, the stellar
temperature determinations are inconsistent with the hot star model.
The line ratios can also be fitted by a combination of a
photoionization model for metal rich HII regions excited by stars of
$\sim$ 38,000K, and the shock excited spectrum of supernova
explosions occurring at the rate measured in the starburst. These
components fit \gal\ with virtually no adjustment of free
parameters. We suggest that many weak-\oa\ LINERs can be explained in
the same manner, rather than by the presence of a mini-AGN or by hot
stars.

\noindent 5.) We have used the infrared data to derive
extinction-independent estimates of the primary boundary conditions
for starburst models of \gal: mass, \nlyc, maximum stellar
temperature, bolometric luminosity, and CO absorption depth.

\noindent 6.) We find that IMF8, proposed by RLRT93 as the best-fitting 
for M~82, also provides a good fit to the properties of \gal\ and
suggests a starburst mass of $6 \times 10^7$~\msun. IMF3, which was
assembled by RLRT93 to represent various representations of the local
IMF, is not satisfactory although it cannot be ruled out as absolutely
as for extreme galaxies like M~82 and NGC 1614.  IMF8 is very similar
to the modified Salpeter IMF that other workers have recently found
gives a good fit in starburst galaxy models. Due to the large
variations or uncertainties in IMFs above 1~\msun, it is no longer
clear whether IMF3 is a meaningful representation of the local stellar
population to be compared with those in starbursts.

\noindent 7.) The model giving the best fit assumes that the star
formation is extended over 20 to 30 million years. A portion of the
supernovae is associated with stars formed near the beginning of the
burst, while the ionizing flux arises from the most recently formed
stars. The rate of star formation must have declined rapidly during
the last roughly 5 million years, to account for the absence of very
hot stars. This model is in agreement with our suggestion that a
HII/weak-\oa\ LINER can arise from a late phase of a starburst.

\acknowledgements

This research has made use of the NASA/IPAC extragalactic database
(NED) which is operated by the Jet Propulsion Laboratory, Caltech,
under contract with the National Aeronautics and Space Administration.
NSO/Kitt Peak FTS data used here were produced by NSF/NOAO.  The
authors would like to thank C. A. Kulesa, J. H. Lacy, K. Luhman,
R. Mauersberger, and H.-W. Rix for useful discussions.  We would like
to thank NSF for support through grants AST91-16442 and AST95-29190.
DMK acknowledges support from NSF grants AST90-20292 (Texas) and
AST94-53354 (Wyoming).  We thank the anonymous referee and our editor,
Steven Willner, for comments which improved this paper.


\clearpage

\begin{deluxetable}{llllll}
\footnotesize
\tablewidth{0pt}
\tablecaption{\gal\ Aperture Photometry}
\tablehead{\colhead{aperture(\arcsec)} & \colhead{$m_J$} & \colhead{$m_H$} & \colhead{$m_K$} & \colhead{$J-H$} & \colhead{$H-K$}}
\startdata
$3$ & $11.55$ & $10.41$ & $9.51$ & $1.14$ & $0.90$ \nl
$6$ & $10.36$ & $9.19$ & $8.44$ & $1.17$ & $0.74$ \nl
$12$ & $9.34$ & $8.25$ & $7.56$ & $1.09$ & $0.69$ \nl
$15$ & $9.05$ & $7.96$ & $7.30$ & $1.09$ & $0.66$ \nl
$2.4\times12$ & $10.41$ & $9.21$ & $8.45$ & $1.20$ & $0.76$ \nl
\label{tbl:phot}
\enddata
\tablecomments{Aperture sizes are diameters of circular apertures
except for the $2\farcs4\times12\arcsec$ aperture which was used to
calibrate the spectra as described in the text.  Typical statistical
uncertainties are 2\% - 5\%.}
\end{deluxetable}

\begin{deluxetable}{ll}
\footnotesize
\tablewidth{0pt}
\tablecaption{Summary of Starburst Parameters}
\tablehead{\colhead{Parameter} & \colhead{Value}}
\startdata
Mass & $\ll 2.1\times10^8$~\msun \nl
\lbol & $1.1\times10^{10}$~\lsun \nl
M$_{K}$ & -20.4 \nl
log \nlyc & $>53.0$ \nl
SNR & $0.03\ {\rm yr}^{-1}$ \nl
CO & 0.20 \nl
\label{tbl:summ}
\enddata
\end{deluxetable}

\begin{deluxetable}{llll}
\footnotesize
\tablewidth{0pt}
\tablecaption{Hydrogen Recombination Line Measurements}
\tablehead{\colhead{} & \colhead{} & \multicolumn{2}{c}{Ratio to \bg}
\\ \cline{3-4} \\
   \colhead{Line} & \colhead{$\lambda_{vac}$(\micron)}
   & \colhead{Observed} & \colhead{Case B}}
\startdata
\pb  & 1.2822 & $1.66\pm0.10$ & 5.58 \nl
Br13 & 1.6114 & $0.10\pm0.03$ & 0.14 \nl
Br12 & 1.6412 & \nodata & 0.18 \nl
Br11 & 1.6811 & $0.12\pm0.03$ & 0.24 \nl
Br10 & 1.7367 & $0.26\pm0.05$ & 0.32 \nl
\bd  & 1.9451 & $0.61\pm0.04$ & 0.65 \nl
\bg  & 2.1661 & $1.00$ & 1.00 \nl
\label{tbl:hlines}
\enddata
\tablecomments{No corrections for extinction have been made.  The flux
in \bg\ is $2.59\pm0.10\times10^{-13}$~\cgs\ in a
2\farcs4$\times$12\arcsec\ aperture.  The Case~B values were derived
from Hummer \& Storey (1987), for $n=10^2$~cm$^{-3}$, $T=5000$~K.}
\end{deluxetable}

\begin{deluxetable}{lll}
\footnotesize
\tablewidth{0pt}
\tablecaption{[\ion{Fe}{2}] Lines}
\tablehead {\colhead{Transition} & \colhead{$\lambda_{vac}$(\micron)}
   & \colhead{Ratio to $\lambda$1.644\micron}}
\startdata
$a^6D_{9/2}-a^4D_{7/2}$ & 1.2570 & $0.68\pm0.04$ \nl
$a^6D_{3/2}-a^4D_{3/2}$ & 1.2791 & $0.04\pm0.01$ \nl
$a^6D_{5/2}-a^4D_{5/2}$ & 1.2947 & $0.13\pm0.03$ \nl
$a^6D_{7/2}-a^4D_{7/2}$ & 1.3209 & $0.18\pm0.02$ \nl
$a^4F_{9/2}-a^4D_{5/2}$ & 1.5339 & $0.13\pm0.03$ \nl
$a^4F_{9/2}-a^4D_{7/2}$ & 1.6439 & 1.00 \nl
$a^4F_{7/2}-a^4D_{5/2}$ & 1.6774 & $0.06\pm0.02$ \nl
\label{tbl:felines}
\enddata
\tablecomments{No correction has been made for extinction.  The flux
in the $\lambda$1.644\micron\ line in a 2\farcs4$\times$12\arcsec\
aperture is $3.56\pm0.18\times10^{-13}$~\cgs.}
\end{deluxetable}

\begin{deluxetable}{lll}
\footnotesize
\tablewidth{0pt}
\tablecaption{He Lines}
\tablehead {\colhead{Transition} & \colhead{$\lambda_{vac}$(\micron)} &
   \colhead{$10^{-14}$~\cgs}}
\startdata
$3P-4D$ & 1.7007 & $<1.0$ \nl
$2S-2P$ & 2.0587 & $9.38\pm1.1$ \nl
\label{tbl:helines}
\enddata
\tablecomments{These fluxes were extracted from a
2\farcs4$\times$12\arcsec\ aperture and are uncorrected for
extinction.}
\end{deluxetable}

\begin{deluxetable}{ll}
\footnotesize
\tablewidth{0pt}
\tablecaption{Velocity gradients of spectral features}
\tablehead{\colhead{Feature} & \colhead{gradient (\kms/\arcsec)}}
\startdata
\bg & $7.4\pm0.4$ \nl
\h2 & $7.7\pm0.4$ \nl
CO(2,1) & $7.5\pm0.9$ \nl
\label{tbl:velgrad}
\enddata
\tablecomments{These numbers are not corrected for inclination.}
\end{deluxetable}

\begin{deluxetable}{lll}
\footnotesize
\tablewidth{0pt}
\tablecaption{Absorption features}
\tablehead{\colhead{Species} & \colhead{$\lambda_{vac}$(\micron)}
   & \colhead {W$_\lambda(\rm\AA)$}}
\startdata
MgI            & 1.4882 & 1.6 \nl
MgI            & 1.5048 & 5.5 \nl
$^{12}$CO(3,0) & 1.5582 & 4.0 \nl
MgI            & 1.5770 & 1.1 \nl
$^{12}$CO(4,1) & 1.5780 & 5.6 \nl
FeI            & 1.5823 & 3.4 \nl
SiI            & 1.5893 & 4.6 \nl
SiI            & 1.5964 & 3.3 \nl
$^{12}$CO(5,2) & 1.5982 & 3.6 \nl
$^{12}$CO(6,3) & 1.6189 & 7.0 \nl
$^{12}$CO(8,5) & 1.6618 & 2.2 \nl
AlI            & 1.6755 & 3.1 \nl
$^{12}$CO(9,6) & 1.6840 & 4.3 \nl
$^{12}$CO(10,7) & 1.7067 & 2.5 \nl
MgI            & 1.7146 & 2.3 \nl
FeI            & 1.7307 & 1.6 \nl
SiI            & 1.7332 & 3.0 \nl
FeI            & 2.0704 & 1.0 \nl
MgI            & 2.1066 & 1.2 \nl
AlI            & 2.1170 & 1.4 \nl
SiI            & 2.1360 & 1.4 \nl
NaI            & 2.207 & 3.2 \nl
FeI            & 2.2263 & 2.0 \nl
CaI            & 2.2631 & 3.3 \nl
CaI            & 2.2657 & 1.8 \nl
MgI            & 2.2814 & 1.1 \nl
$^{12}$CO(2,0) & 2.2935 & 13.1 \nl
$^{12}$CO(3,1) & 2.3227 & 16.4 \nl
$^{13}$CO(2,0) & 2.3448 & 10.8 \nl
$^{12}$CO(4,2) & 2.3525 & 16.4 \nl
$^{13}$CO(3,1) & 2.3739 & 13.0 \nl
$^{12}$CO(5,3) & 2.3830 & 17.1 \nl
\label{tbl:abs}
\enddata
\end{deluxetable}

\begin{deluxetable}{lllllll}
\footnotesize
\tablewidth{0pt}
\tablecaption{\h2\ Line Fluxes}
\tablehead{
   \colhead{} & \colhead{} & \multicolumn{5}{c}{Ratio to (1,0)S(1)} \\
   \cline{3-7} \\
   \colhead{Transition} & \colhead{$\lambda_{vac}$(\micron)}
   & \colhead {Obs.} & \colhead{Dered.} & \colhead{Fluor.}
   & \colhead {Therm.} & \colhead {Mix}}
\startdata
(5,3)Q(1) & 1.4929 & $<0.1$ & $<0.18$ & 0.43 & $1.0\times10^{-4}$ &
0.11 \nl
(4,2)O(3) & 1.5099 & $<0.1$ & $<0.18$ & 0.42 & $7\times10^{-4}$ &
0.11 \nl
(6,4)Q(1) & 1.6015 & $<0.15$ & $<0.24$ & 0.33 & $1.3\times10^{-4}$ &
0.08 \nl
(5,3)O(3) & 1.6135 & $<0.15$ & $<0.23$ & 0.38 & $9\times10^{-5}$ &
0.10 \nl
(6,4)O(3) & 1.7326 & $<0.15$ & $<0.20$ & 0.31 & $1\times10^{-5}$ &
0.08 \nl
(1,0)S(3) & 1.9576 & $0.89\pm0.08$ & $0.99\pm0.09$ & 0.67 & 1.02 &
0.93 \nl
(2,1)S(4) & 2.0041 & $0.18\pm0.02$ & $0.19\pm0.02$ & 0.12 & 0.02 &
0.05 \nl
(1,0)S(2) & 2.0338 & $0.36\pm0.05$ & $0.38\pm0.05$ & 0.50 & 0.38 &
0.41 \nl
(2,1)S(3) & 2.0735 & $0.26\pm0.06$ & $0.27\pm0.06$ & 0.35 & 0.08 &
0.15 \nl
(1,0)S(1) & 2.1218 & $1.00       $ & 1.00          & 1.00 & 1.00 &
1.00 \nl
(2,1)S(2) & 2.1542 & $0.20\pm0.04$ & $0.20\pm0.04$ & 0.28 & 0.04 &
0.09 \nl
(1,0)S(0) & 2.2233 & $0.30\pm0.03$ & $0.28\pm0.03$ & 0.46 & 0.21 &
0.27 \nl
(2,1)S(1) & 2.2477 & $0.21\pm0.02$ & $0.20\pm0.02$ & 0.56 & 0.08 &
0.21 \nl
(2,1)S(0) & 2.3556 & $<0.1$ & $<0.09$ & 0.26 & 0.02 &
0.08 \nl
(3,2)S(1) & 2.3865 & $<0.1$ & $<0.09$ & 0.29 &
$5.8\times10^{-3}$ & 0.08 \nl
(1,0)Q(1) & 2.4066 & $0.79\pm0.16$ & $0.68\pm0.14$ & 0.99 & 0.70 & 0.77 \nl
(1,0)Q(2) & 2.4134 & $0.34\pm0.07$ & $0.29\pm0.06$ & 0.51 & 0.23 & 0.31 \nl
(1,0)Q(3) & 2.4237 & $0.75\pm0.15$ & $0.64\pm0.13$ & 0.70 & 0.70 & 0.70 \nl
(1,0)Q(4) & 2.4375 & $0.20\pm0.04$ & $0.17\pm0.03$ & 0.28 & 0.21 & 0.23 \nl
\label{tbl:h2lines}
\enddata
\tablecomments{The flux in the (1,0)S(1) line is
$1.21\pm0.07\times10^{-13}$~\cgs\ (uncorrected for
extinction) in a 2\farcs4$\times$12\arcsec\ aperture.}
\end{deluxetable}

\begin{deluxetable}{ccr}
\footnotesize
\tablewidth{0pt}
\tablecaption{\h2\ rotational excitation temperatures}
\tablehead{\colhead{Vibrational} & \colhead{Rotational}
   & \colhead{Rotational Excitation} \\ \colhead{Level}
   & \colhead{Levels} & \colhead {Temperature (K)}}
\startdata
1 & 0,2 & $1350\pm229$ \nl
  & 1,3 & $1921\pm175$ \nl
2 & 1,3 & $3528\pm860$ \nl
\label{tbl:rottemp}
\enddata
\end{deluxetable}

\begin{deluxetable}{llll}
\footnotesize
\tablewidth{0pt}
\tablecaption{Predicted optical line ratios}
\tablehead{\colhead{ratio} & \colhead{predicted} & \colhead {observed}
  & \colhead{ref.}}
\startdata
\oa/H$\alpha$     &   0.026    &      0.044   &       1 \nl
[\ion{N}{2}]/H$\alpha$  &   0.78     &      0.78    &       1 \nl
[\ion{S}{2}]/H$\alpha$  &   0.45     &      0.40    &       1 \nl
\oc/H$\beta$  &   0.49     &      0.47    &       2 \nl
\label{tbl:nebular}
\enddata
\tablecomments{(1) Armus, Heckman, \& Miley 1989; (2) Tadhunter \etal\
1993.}
\end{deluxetable}


\clearpage

\clearpage

\clearpage

\clearpage

\clearpage

\clearpage

\clearpage

\clearpage

\clearpage

\clearpage

\clearpage

\clearpage

\begin{figure}
\plotone{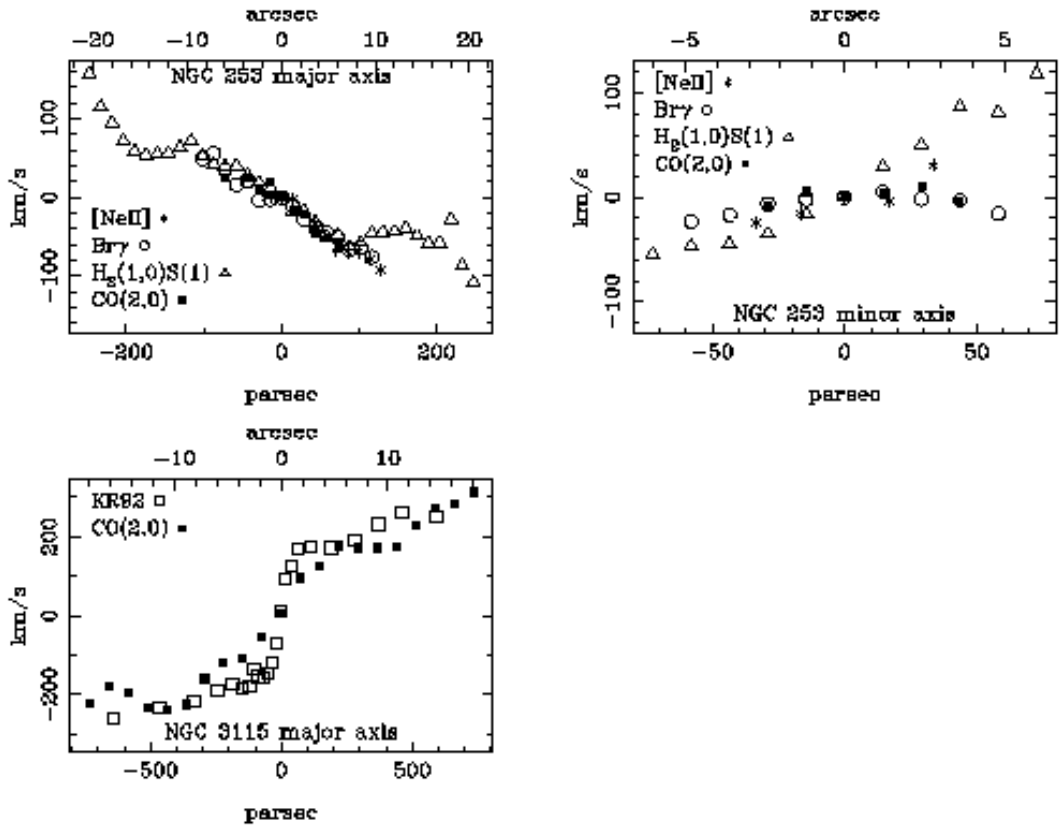}
\caption{Rotation curves for \gal\ and NGC~3115.  NE is on the right
and SW to the left in the \gal\ major-axis plot while SE is on the
right and NW to the left in the minor-axis plot.  The linear
dimensions have been determined by assuming distances of 2.5~Mpc and
8.4~Mpc, respectively, for \gal\ and NGC~3115.}
\label{fig:rotcurve}
\end{figure}\clearpage

\clearpage

\clearpage

\end{document}